\documentclass{MICE}
\usepackage{lineno}
\usepackage{bibentry}
\nobibliography*
\usepackage{times}
\usepackage{bm}         % bold maths
\usepackage{amsmath,amssymb,bm,slashed} % need for subequations
\usepackage{amssymb}    % extra math symbols
\usepackage{graphicx}   % need for figures
\usepackage{verbatim}   % useful for program listings
\usepackage{color}      % use if color is used in text
\usepackage{subfigure}  % use for side-by-side figures
\usepackage{hyperref}   % use for hypertext links, including those to external documents and URLs
\usepackage{enumerate}
\usepackage{lineno}
\usepackage{textcomp}

\usepackage{fmtcount}   % Allow counters to be printed with padding

\usepackage[square,comma,numbers,sort&compress]{natbib}

\def\parenbar{\mathpalette\p@renb@r}
\def\p@renb@r#1#2{\vbox{%
  \ifx#1\scriptscriptstyle \dimen@.7em\dimen@ii.2em\else
  \ifx#1\scriptstyle \dimen@.8em\dimen@ii.25em\else
  \dimen@1em\dimen@ii.4em\fi\fi \offinterlineskip
  \ialign{\hfill##\hfill\cr
    \vbox{\hrule width\dimen@ii}\cr
    \noalign{\vskip-.3ex}%
    \hbox to\dimen@{$\mathchar300\hfil\mathchar301$}\cr
    \noalign{\vskip-.3ex}%
    $#1#2$\cr}}}

\newenvironment{alphafootnotes}
  {\par\edef\savedfootnotenumber{\number\value{footnote}}
   \renewcommand{\thefootnote}{\fnsymbol{footnote}}
   \setcounter{footnote}{0}}
  {\par\setcounter{footnote}{\savedfootnotenumber}}

\begin{document}
% --------------------------------------- %
% --------------TOP MATTER--------------- %
% --------------------------------------- %
\thispagestyle{empty}

\begin{tabular}{p{0.175\textwidth} p{0.5\textwidth} p{0.225\textwidth}}
  \hspace{-0.8cm}\leftline{RAL-P-2015-008}                                 &
  \centering{Muon Ionization Cooling Experiment}                  &
  \rightline{\today} 
\end{tabular}
\vspace{-1.0cm}\\
\rule{\textwidth}{0.43pt}

\renewcommand{\thefootnote}{\ifcase\value{footnote}\or$\dagger$\or*\or
$\ddagger$\or\#\or$\dagger\dagger$\or**\or$\ddagger\ddagger$\or \#\#\or $\infty$\fi}
\begin{center}
  {\bf
    {\LARGE Electron-Muon Ranger:\\performance in the MICE Muon Beam} \\
  }
  \vspace{0.2cm}
  The MICE collaboration\footnote{Authors are listed at the end of this paper.} \\
  \vspace{-0.0cm}
\end{center}
\renewcommand{\thefootnote}{\arabic{footnote}}
\setcounter{footnote}{0}

\makeatletter

\parindent 10pt
\pagestyle{plain}
\pagenumbering{arabic}                   
\setcounter{page}{1}

% --------------------------------------- %
% ---------------ABSTRACT---------------- %
% --------------------------------------- %
\begin{quotation}
\noindent
The Muon Ionization Cooling Experiment (MICE) will perform a detailed study of ionization cooling to evaluate the feasibility of the technique. To carry out this program, MICE requires an efficient particle-identification (PID) system to identify muons. The Electron-Muon Ranger (EMR) is a fully-active tracking-calorimeter that forms part of the PID system and tags muons that traverse the cooling channel without decaying. The detector is capable of identifying electrons with an efficiency of 98.6\%, providing a purity for the MICE beam that exceeds 99.8\%. The EMR also proved to be a powerful tool for the reconstruction of muon momenta in the range 100--280\,MeV/$c$.
\end{quotation}

% --------------------------------------- %
% -------------INTRODUCTION-------------- %
% --------------------------------------- %
\section{Introduction}
\label{sec:introduction}

Intense muon sources are required for a future Neutrino Factory or Muon Collider~\cite{nf, mucol}.  At production, muons occupy a large phase-space volume (emittance), which makes them difficult to accelerate and store.  Therefore, the emittance of the muon beams must be reduced, i.e the muons must be ``cooled'', to maximise the muon flux delivered to the accelerator. Conventional cooling techniques applied to muon beams~\cite{Parkhmochuck} would leave too few muons to be accelerated since the muon lifetime is short ($\tau_\mu\sim2.2\,\mu$s). Simulations indicate that the ionization-cooling effect builds quickly enough to deliver the flux and emittance required by the Neutrino Factory and the Muon Collider~\cite{nfISS, Palmer:1996gs}. The MICE collaboration will study ionization cooling in detail to demonstrate the feasibility of the technique~\cite{Gregoire:2003nh}.

Ionization cooling proceeds by passing a beam of muons through a low-$Z$ material~\cite{Neuffer:1983jr}.  The beam loses energy by ionizing the material, reducing its total momentum.  Longitudinal momentum is restored by accelerating cavities. The net effect is to reduce the divergence of the beam and the transverse phase-space the beam occupies.  The rate of change of the normalised 2D emittance may be approximated by~\cite{Neuffer:multitev}:
\begin{equation}
\frac{d\varepsilon_{N}}{ds} \simeq -\frac{\varepsilon_{N}}{\beta^{2} E_{\mu}} \left|\frac{dE_{\mu}}{ds}\right|
+ \frac{\beta_{\perp} \left( 0.014 \right) ^{2}}{2\beta^{3}E_{\mu}m_{\mu}X_{0}}\, ;
\label{Eq:CoolingFormula}
\end{equation}
where $\beta=v/c$,  $E_{\mu}$,  $m_{\mu}$ are the muon velocity, energy and mass respectively.  The rate of change of emittance depends on the properties of the absorber and the beam.  Cooling is large when the initial emittance of the beam, $\varepsilon_{N}$, and stopping power of the absorber, $\left \langle dE_\mu/ds\right \rangle$, are large.  The effect of heating by multiple Coulomb scattering is reduced if the radiation length of the absorber, $X_{0}$, is large and the transverse betatron function, $\beta_{\perp}$, of the beam at the absorber is small. Optimum cooling is achieved with low-$Z$ absorbers, such as liquid hydrogen or lithium hydride, and with solenoidal beam-focussing.

The muon beams at the front-end of a Neutrino Factory or Muon Collider are expected to be similar, with a large transverse normalised emittance of $\varepsilon_{N}\approx 12$--$20$\,$\pi$\,mm-rad and a momentum spread of $\sim 20$\,MeV/$c$. The emittance must be reduced to 2--5\,$\pi$\,mm-rad for the Neutrino Factory, with further reduction to 0.008\,$\pi$\,mm-rad required for a Muon Collider~\cite{higgsFactory}. The Muon Ionization Cooling Experiment (MICE)~\cite{mice:web} collaboration intends to demonstrate the feasibility of an ionization-cooling cell suitable for cooling muon beams at a Neutrino Factory. An accurate measurement of the degree of cooling achieved depends on selecting a pure sample of muons by rejecting their parent particles or decay products. The Electron-Muon Ranger (EMR)~\cite{Asfandiyarov:2013wva} is a key component of the particle identification system and its performance in the MICE Muon Beam (MMB) is presented here.

\subsection{The Muon Ionization Cooling Experiment}\label{sec:MICE}

Since energy loss by ionization and multiple Coulomb scattering are momentum dependent, the ionization-cooling effect is momentum dependent. MICE will measure the performance of its cooling cell for muon beams with central momenta in the range 140--240\,MeV/$c$ and a momentum spread of $\sim 20$\,MeV/$c$, as would be the case in a Neutrino Factory. This is achieved using a ``super focus-focus'' lattice cell~\cite{SFOFO} which is capable of producing a range of $\beta_{\perp}$ at the position of the absorber.

\begin{figure}[tbp]
\centering
\includegraphics[height=.8\textwidth, angle=90]{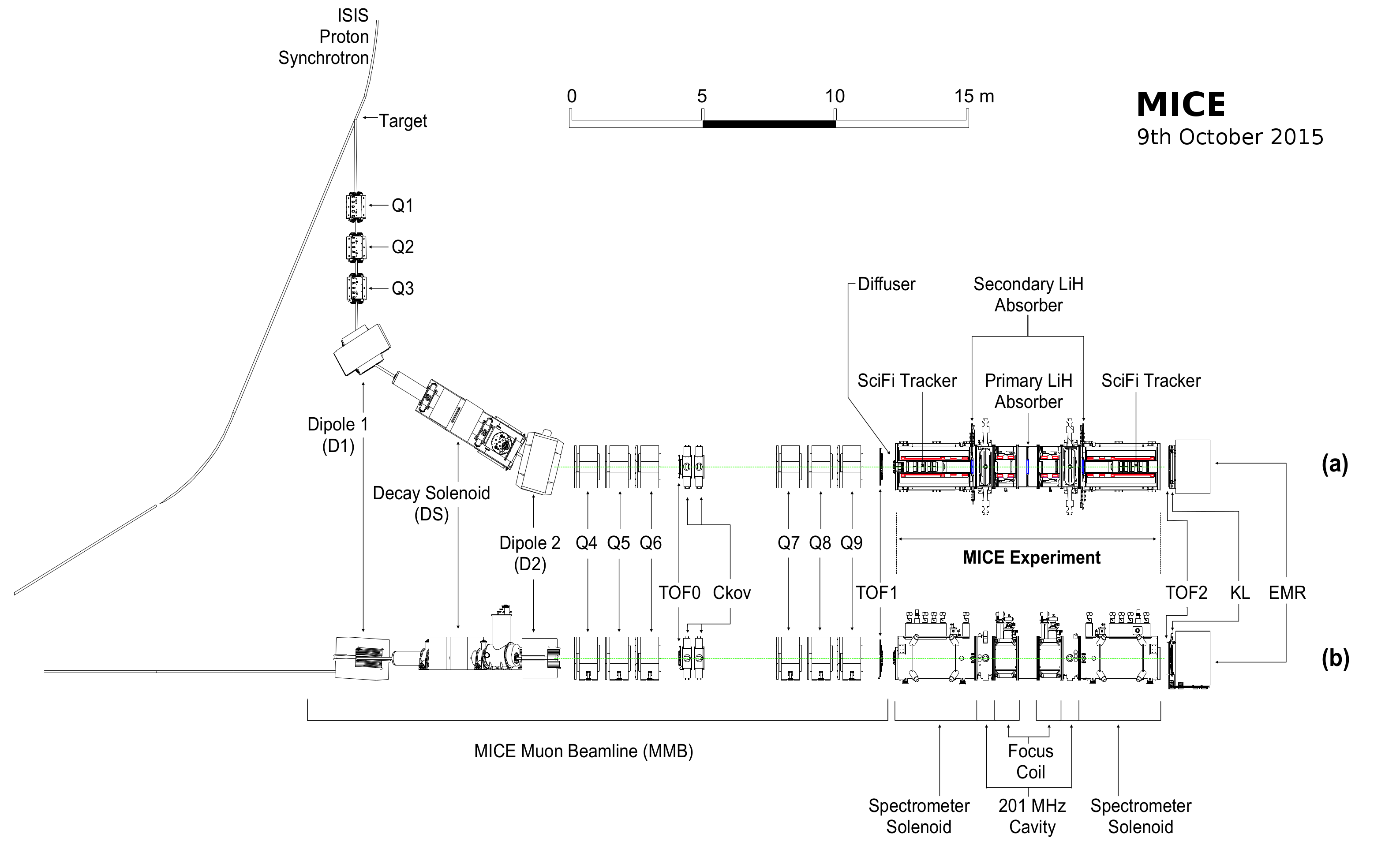}
\caption{(a) Cross-sectional and (b) side views of the MICE Muon Beam Line (MMB) and the MICE experiment. The Electron-Muon Ranger immediately precedes the beam stop and is the final component of the experiment. The MICE magnet channel (from upstream spectrometer solenoid up to and including the downstream sprectrometer solenoid) was not present when the data reported here was taken.}
\label{fig:BeamLine}
\end{figure}

A schematic of the MICE Muon Beam (MMB) and the MICE experiment is shown in figure~\ref{fig:BeamLine} and described in detail in~\cite{BeamlinePaper, IonizationCoolingDemoPaper}. The MMB operates on the ISIS proton synchrotron~\cite{isis:web} at the Rutherford Appleton Laboratory. A titanium target~\cite{target} samples the ISIS proton beam, producing pions.  The pions are transported by the upstream quadrupoles, Q1--3, and are momentum-selected at the first dipole, D1. The high field present in the Decay Solenoid (DS) increases the pion path-length with the result that the bulk of pion decays occur as the beam passes through the DS. The second dipole, D2, is used to momentum-select a `muon' beam with high purity or a `calibration' beam containing a mix of muons, pions and electrons.  The resultant beam is transported through two quadrupole triplets, Q4--6 and Q7--9, time-of-flight counters~\cite{TOFref}, TOF0 and TOF1, and Cherenkov detectors, Ckov~\cite{MICE_PID}, to the cooling cell. The beam emittance is inflated as it passes through the variable-thickness brass and tungsten `diffuser' and is measured in the upstream spectrometer solenoid using a scintillating-fibre tracker.  The beam then passes through low-$Z$ absorbers and RF cavities prior to being remeasured in the downstream tracker.  Upon exiting the cooling cell the beam is incident upon the final time-of-flight detector, TOF2~\cite{Bertoni:tof2}, a pre-shower detector, KL~\cite{MICE_PID}, and the Electron-Muon Ranger, EMR~\cite{Asfandiyarov:1966952}. 

The particle identification system consists of the TOF0, TOF1 and TOF2, Ckov, KL and EMR detectors, which can conceptually be split in two; TOF0--1 and Ckov identifying particle species prior to the cooling cell and TOF2, KL and EMR identifying species after the cooling cell.  In combination with time-of-flight information, the EMR is used to distinguish between muons that have successfully traversed the full cooling cell from those that have decayed en-route.  This aids in the measurement of beam transmission through the cooling cell, as well as reducing the uncertainty on the emittance measurement.

% --------------------------------------- %
% ------------EMR DESCRIPTION------------ %
% --------------------------------------- %
\section{The Electron-Muon Ranger in the MICE Muon Beam}
\label{Sect:Method}

The primary purpose of the EMR is to distinguish between muons and their decay products, identifying muons that have crossed the entire cooling channel~\cite{Asfandiyarov:1966952}. This allows for the selection of a muon beam with a contamination below 1\%.  The EMR is a tracking calorimeter, consisting of multiple, orthogonal, layers of triangular scintillating bars arranged in planes. This construction allows several discriminating parameters to be determined that may be used to identify particle species at the EMR.

\subsection{EMR construction}\label{SubSect:Construction}

\begin{figure}[tbp]
 \centering
 \hspace*{\fill}
 \begin{minipage}[t]{.45\textwidth}
 \includegraphics[width=\columnwidth]{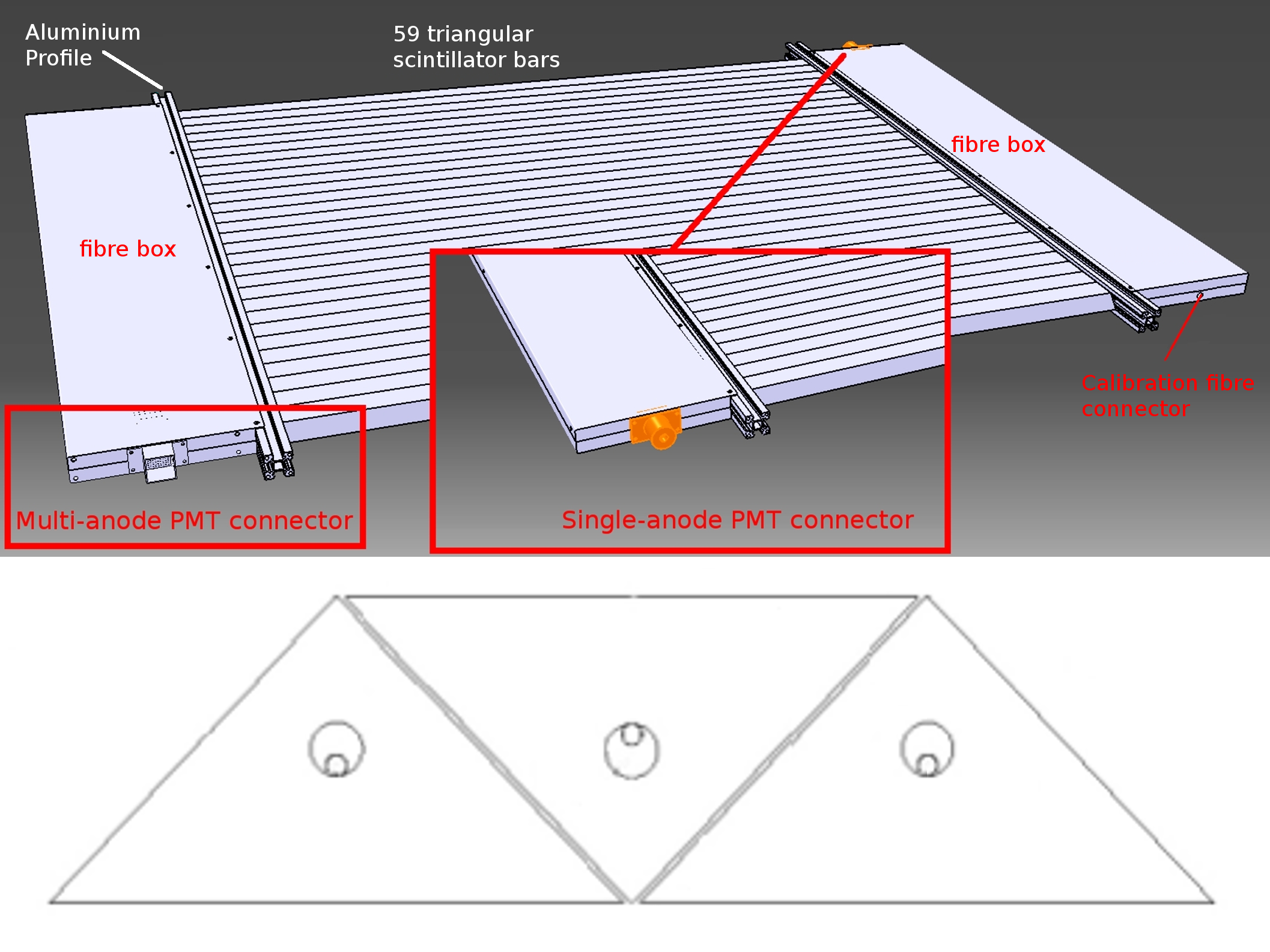}
 \caption{CAD drawing of one EMR plane (top) and cross section of 3 bars and their threaded WLS fibres (bottom).\label{Fig:EMRPlane}}
 \end{minipage}
 \hspace*{\fill}
 \begin{minipage}[t]{.45\textwidth}
 \includegraphics[width=\columnwidth]{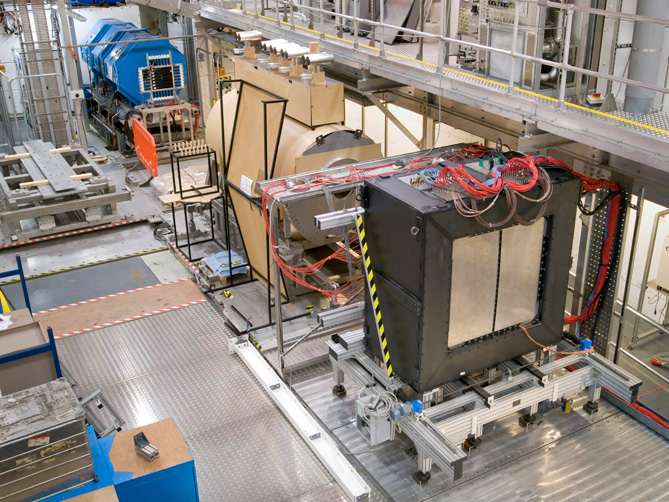}
 \caption{The EMR detector installed in the MICE Hall at the end of the preliminary (`Step I') beam line. The photograph also shows Q7--9 and TOF1 at the top half of the frame. A model of one spectrometer solenoid  was positioned in the hall to allow installation procedures to be developed.\label{Fig:EMRInHall}}
 \end{minipage}
 \hspace*{\fill}
\end{figure}

Figure~\ref{Fig:EMRPlane} shows one plane of the EMR, consisting of 59, tessellated, triangular scintillator bars covering an area of 1.21\,m$^{2}$.  Each plane is rotated through 90$^\circ$ with respect to the previous one, such that a pair of planes defines a horizontal and vertical $(x,y)$ interaction coordinate.  Light produced by a particle interaction is collected by the wavelength shifting (WLS) fibre within each bar and is read out at both ends.  On one side of the plane, the 59 fibres are bundled together and brought to a conventional single-anode photo-multiplier tube (SAPMT) and on the other side the fibres are individually brought to one of the pixels on a 64-channel multi-anode photo-multiplier tube (MAPMT). A dedicated front-end board reads out and digitizes the MAPMT signals individually by means of fast shapers and discriminators.

Figure~\ref{Fig:EMRInHall} shows the full detector, consisting of 48 planes, installed in the MICE Hall at the end of the preliminary (`Step I') beam line~\cite{BeamlinePaper}.  It is enclosed on all sides to ensure a light-tight environment and is magnetically shielded from the cooling cell.  A complete description of the design and fabrication of the EMR can be found in~\cite{Asfandiyarov:1966952}.

\subsection{Data taking}

In September 2013, the EMR was placed after TOF2 and KL, $\sim10\,$m from TOF1, and exposed to a variety of beams , prior to the installation of the cooling channel. Both `muon' and `calibration' beams over a range of momenta were used to illuminate the detector. These data were used to verify its ability to separate muons from their decay products.

The MMB can be tuned to enhance the production of muons (the `muon' beam), pions or electrons (`calibration' beams).  All beam configurations contain all three particle species, albeit in differing proportions.  Beams are defined according to the preferred particle species, $e,\mu,\pi$ and the momentum selected (in the range 100--400\,MeV/$c$) by the D2 magnet, $p_{\mathrm{D2}}$, for the preferred particle type.

\begin{figure}[tbp]
 \centering
 \begin{minipage}[t]{.45\textwidth}
  \includegraphics[width=\textwidth]{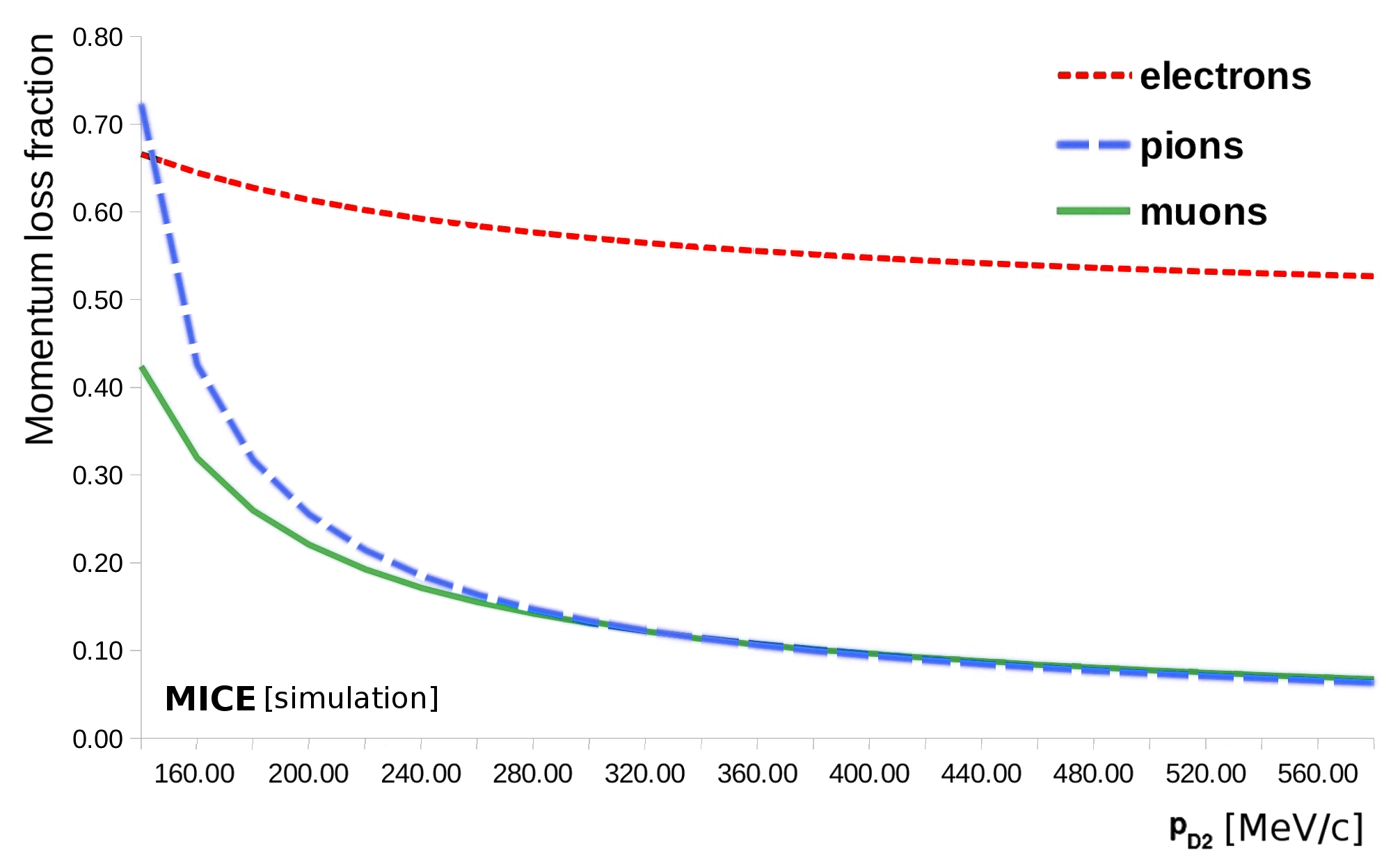}
  \caption{Momentum loss fraction of particles travelling from D2 to the EMR, compared to their momentum at D2 ($p_\mathrm{D2}$).\label{Fig:MomentumLossFraction}}
 \end{minipage}
 \hspace*{\fill}
 \begin{minipage}[t]{.45\textwidth}
  \includegraphics[width=\textwidth]{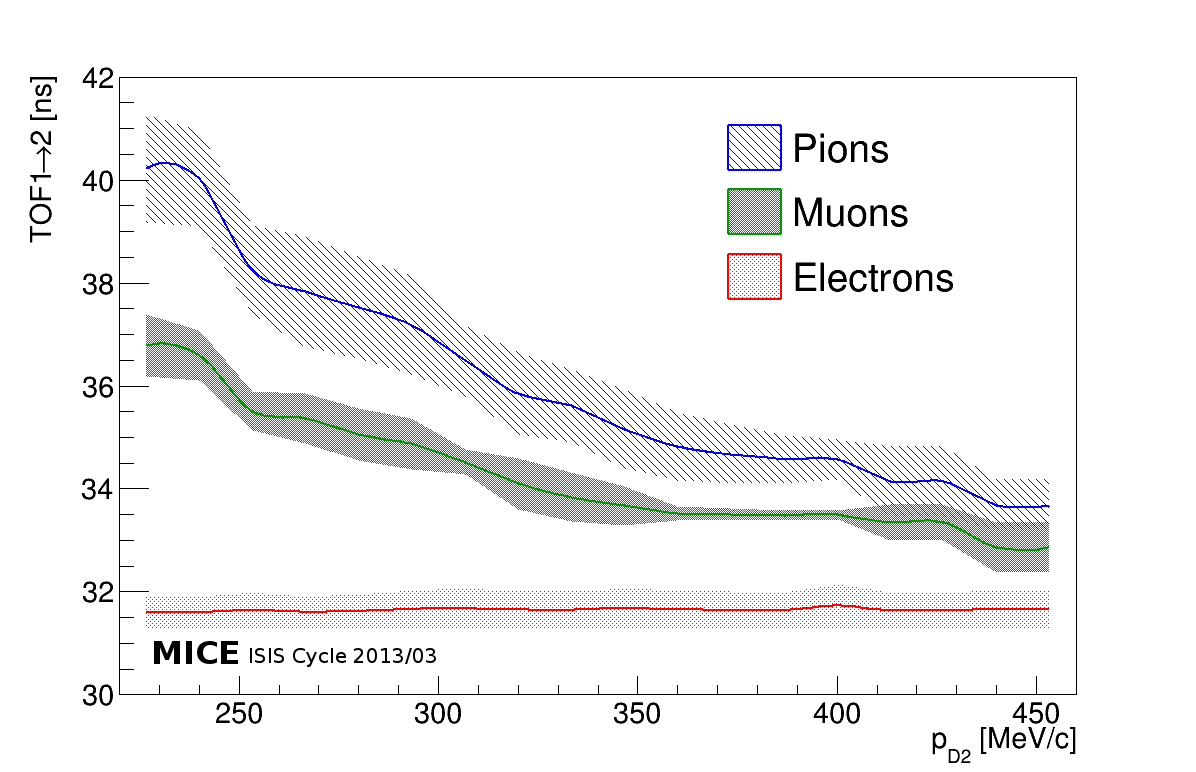}
  \caption{Time-of-flight between the TOF1 and TOF2 detectors (TOF1$\rightarrow$2) for a `calibration' beam selected with an array of momenta at D2 ($p_\mathrm{D2}$).\label{Fig:TOF}}
 \end{minipage}
\end{figure}

Particles travelling from D2 to the EMR cross $\sim$~22.7\,m of air as well as the remaining PID detectors (TOFs, Ckov, KL), losing momentum due to ionization.  This is expressed as a momentum-loss fraction in figure~\ref{Fig:MomentumLossFraction}, i.e. the fraction of momentum lost by a particle given its momentum selected at D2.  As the EMR must identify muons from electrons over a range of momenta, the beam momentum was scanned from 230--450\,MeV/$c$ at D2.  This corresponds to `muon' and `pion' beam momenta in the range 200--420\,MeV/$c$, with `electron' beam momentum range of 100--250\,MeV/$c$.

\subsection{Particle identification by time-of-flight}\label{sec:tof_characterization}

The three time-of-flight detectors, TOF0, TOF1 and TOF2, can be used to measure particle momentum, given a particle-species hypothesis.  Calibration beams exhibit a three-peak pattern, corresponding to electrons, muons and pions in order of increasing time-of-flight as shown  in figure~\ref{Fig:TOF} for an array of D2 settings. The time-of-flight for a particular particle is used to define a particle-type hypothesis and to determine the corresponding momentum via $p/E=s/t$, where $s$ is the path length between TOF$i$ and TOF$j$ and $t=t_{\mathrm{TOF}i} - t_{\mathrm{TOF}j}$.  

TOF0 and TOF1 are separated by 7.8\,m, whereas TOF1 and TOF2 are separated by 9.4\,m. The air between the last two hodoscopes causes a minimum ionizing muon to lose approximately 2\,MeV/$c$, i.e. $\sim1$\% of its momentum. The calculation of the probability of a particle being classed as an electron, muon or pion using time-of-flight, for later assessment by the EMR, is based on the assumption that momentum remains approximately constant and that the path length, $s$, is given by the orthogonal distance between the two TOF planes in question.  A combined probability from the three TOFs (i.e. TOF0$\rightarrow$1, TOF0$\rightarrow$2, TOF1$\rightarrow$2) of greater than 90\% for a particle to be a muon, leads to that particle being accepted as part of the muon sample, and its momentum to be calculated from its time-of-flight TOF1$\rightarrow$2 using the appropriate particle mass. Similarly, a combined probability greater than 90\% for a particle to be an electron, leads to that particle being accepted as part of the electron sample. Only the particles that were successfully tagged were used to study the performance of the EMR.

\subsection{Momentum loss in TOF2 and the KL}

\begin{figure}[tbp]
 \centering
 \begin{minipage}[t]{.5\textwidth}
  \centering
  \includegraphics[width=\textwidth]{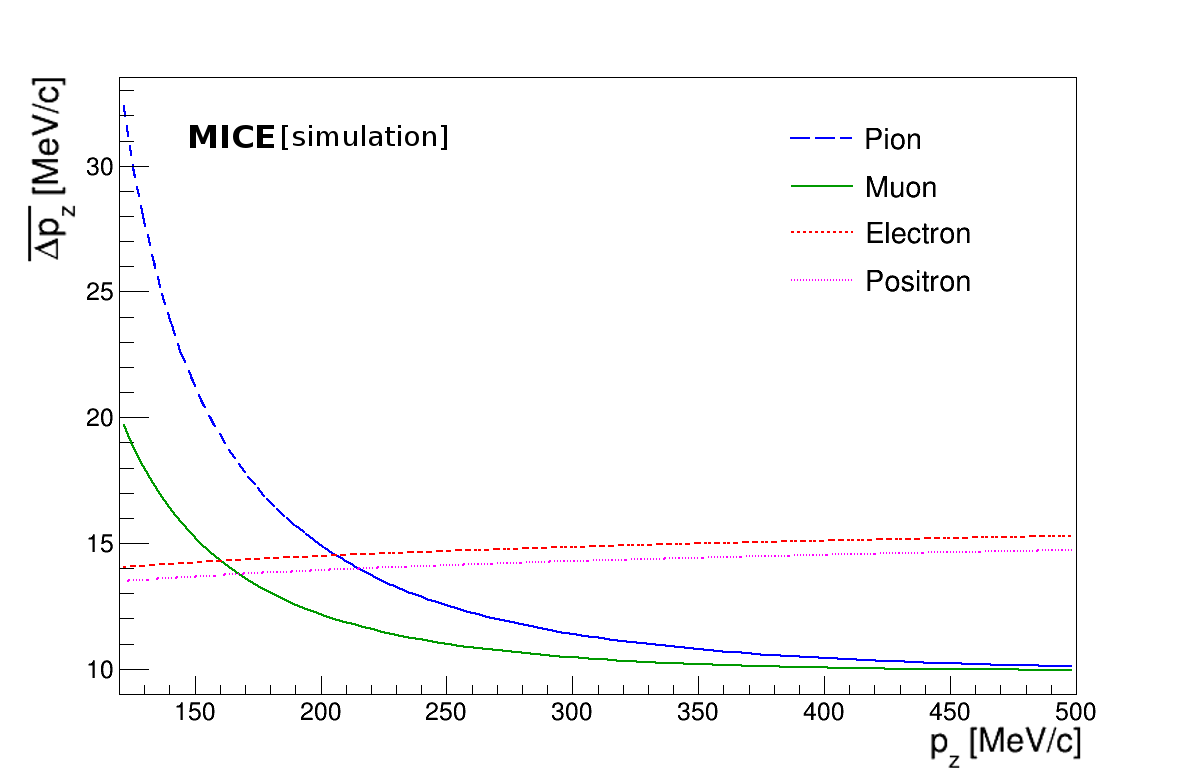}
  \caption{Continuous Slowing Down Approximation of the mean momentum loss ($\overline{\Delta p_z}$) of particles crossing TOF2 with impinging momentum $p_z$.\label{TOFMomentumLoss}}
 \end{minipage}
 \hspace*{\fill}
 \begin{minipage}[t]{.45\textwidth}
  \centering
  \includegraphics[width=\textwidth]{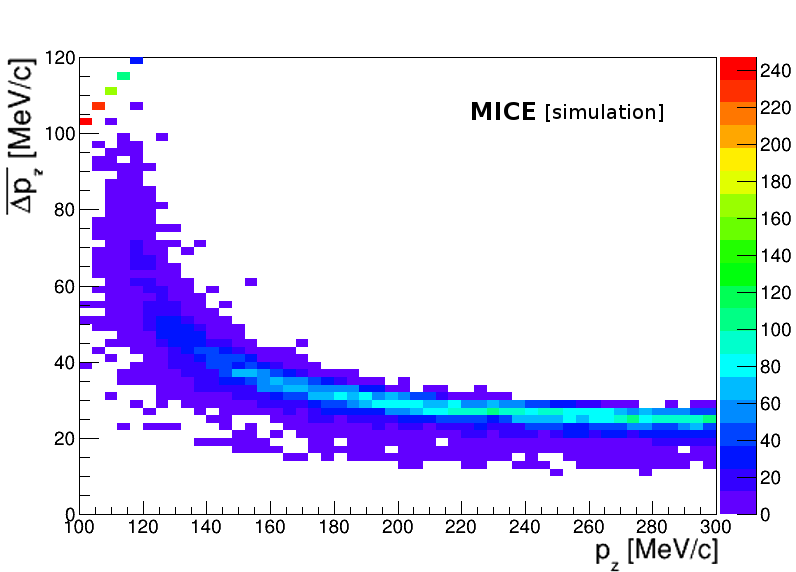}
  \caption{Continuous Slowing Down Approximation of the mean momentum loss ($\overline{\Delta p_z}$) of muons crossing the KL with impinging momentum $p_z$.\label{KLMomentumLoss}}
 \end{minipage}
\end{figure}

Before entering the EMR, particles pass through the final TOF station, TOF2, and the pre-shower detector, KL. The energy lost by particles in TOF2 is calculated using the Continuous Slowing Down Approximation (CSDA)~\cite{carron2006introduction}, where the rate of energy loss at every point along its track is assumed to be the same as the total stopping power.  TOF2 is composed of two, 2.54\,cm thick, planes of polyvinyltoluene (PVT) constituting 0.1\,$X_{0}$.  On average, minimum ionizing muons lose $\sim$\,10\,MeV/$c$, whereas electrons lose $\sim$\,15\,MeV/$c$ as they are ultrarelativistic ($\beta\gamma>100$) in all cases (figure~\ref{TOFMomentumLoss}).

The KL is a sampling calorimeter composed of extruded Pb foils and scintillating fibres in a 1\,:\,2 volume ratio.  On average this corresponds to 2.5\,$X_{0}$.  Electrons shower in the KL before exiting the detector, which makes estimating the energy lost difficult.  Muons cross without showering, but still deposit a significant portion of their energy in the KL.  Due to the complexity of the KL detector, a simplified CSDA approximation was developed to estimate momentum loss in the KL (figure~\ref{KLMomentumLoss}). Minimum ionizing muons lose, on average, $28\pm3$\,MeV/$c$ in the KL.  

\subsection{Events in the EMR}
\label{Sect:EMRPerf}

The EMR is put immediately behind the KL. It is an advantage when using it to identify electron-induced shower since the KL acts as a pre-shower detector.

To characterise the performance of the EMR, particles were first identified with the TOF detectors as described in section~\ref{sec:tof_characterization}. The particle events included in this analysis consist of a single space-point (X- and Y-view hits) in each TOF and the successful assignment of a particle species. These space points were matched to the relevant EMR hits through their temporal proximity.

The coordinates of a hit are taken to be those of the centre of gravity of the triangular section of the corresponding bar, which is located one third of its height up from its base and bisecting its width. The EMR planes (defined in section~\ref{SubSect:Construction}) are placed perpendicular to the beam and their numbering therefore provides the $z$ coordinate. The bar number within that plane determines the second coordinate, $q$, i.e $x$ is the coordinate provided by the X planes and $y$ the coordinate provided by the Y planes.

\begin{figure}[tbp]
 \centering
 \includegraphics[width=.85\textwidth]{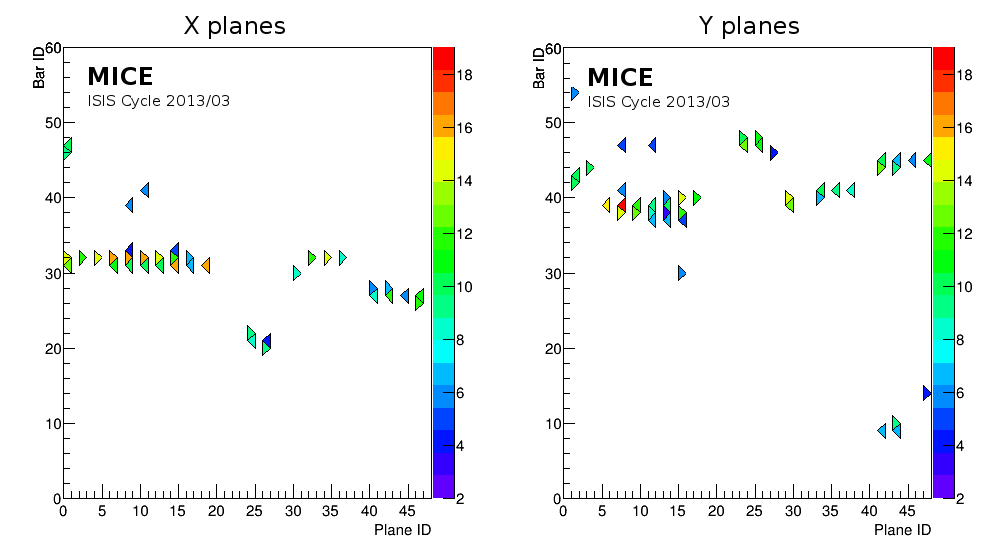}
 \caption[Beam particle event display: electron shower]{EMR event display of the energy deposited by a positron shower ($p_\mathrm{D2}$ = 450\,MeV/$c$) in the two projections. The location of a hit is defined by the plane number (Plane ID, 0--47) and bar number (Bar ID, 1--59) and the energy deposited is represented by the colour code in units of time-over-threshold.}
 \label{fig:singleTrigger_event_display_sp30_tr2}
\end{figure}

An event in the EMR induced by a single positron hitting the KL is represented in the $xz$ and $yz$ projections in figure~\ref{fig:singleTrigger_event_display_sp30_tr2}. The positron originates from an `calibration' beam with a particle momentum of 450\,MeV/$c$ selected at D2. The remaining charged particles in the shower create shallow straight tracks in the first half of the detector, while photons produce deep and detached hits that are scattered far into the scintillating volume. Some planes do not record any hit on the particle\textquotesingle s path, as might be expected in an electromagnetic shower.

\begin{figure}[tbp]
 \centering
 \includegraphics[width=.85\textwidth]{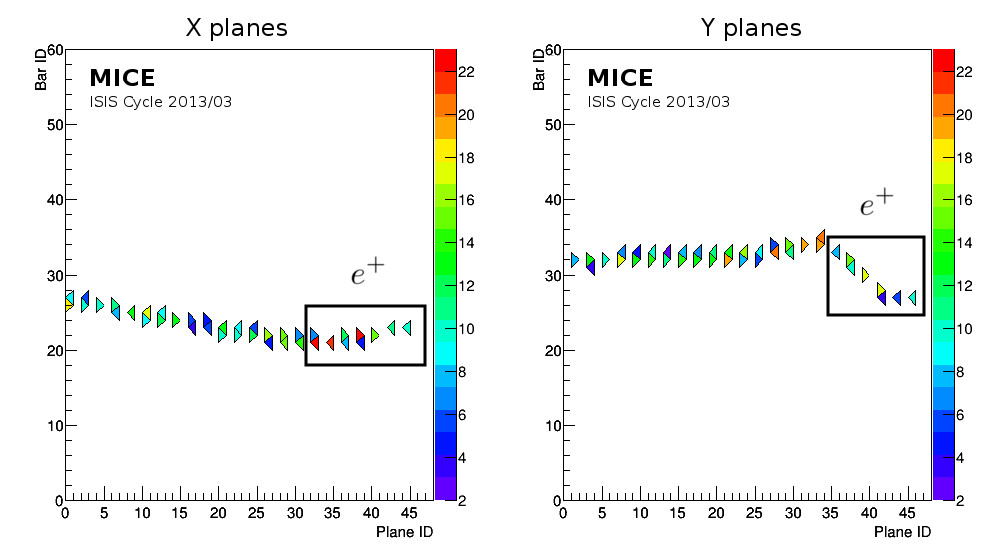}
 \caption[Beam particle event display: muon]{EMR event display of the energy deposited by a $\mu^+$ which decays in the detector volume. The location of a hit is defined by the plane number (Plane ID, 0--47) and bar number (Bar ID, 1--59) and the energy deposited is represented by the colour code in units of time-over-threshold.}
 \label{fig:singleTrigger_event_display_sp33_tr4}
\end{figure}

Figure~\ref{fig:singleTrigger_event_display_sp33_tr4} represents the trace left by a single positive muon stopping in the EMR. The impinging muon stops in the 34$^{\rm th}$ plane where the highest energy deposition is recorded (two bars with a time-over-threshold $\sim$\,20 ADC counts\footnote{The time-over-threshold is defined as the width of the low threshold discriminator signal. This signal width, larger for signal with larger amplitudes, depends on the energy lost by the particle. 1 ADC count corresponds to a signal width of 2.5\,ns.}). The second part of the track, discriminated from the first part using the timing information, is the positron produced by the muon decay.

% --------------------------------------- %
% --------PARTICLE IDENTIFICATION-------- %
% --------------------------------------- %
\section{EMR particle identification variables}

Several variables are defined to reduce the information recorded in a complex event to manageable quantities. Due to the different topology of an electron shower and a muon track, a geometrical approach was chosen. Hits from an electromagnetic shower are widely spread throughout the detector, including far downstream into the fiducial volume, without any visible tracks upstream. Muons exhibit clear straight tracks through the EMR before stopping or exiting the volume. 

Because of the photon-induced hits at large depth, the range of an electron event is not well defined and thus is not a good variable to use to reject electrons. Non-negligible crosstalk~\cite{Drielsma:xt} within the multi-anode PMTs prevents the use of a simple shape analysis as muon events could appear wider, thus potentially increasing the loss of real muons. Alternative longitudinal and transverse evaluations of the event are used to distinguish between the two species: the `plane density' and the `shower spread', respectively.

\subsection{Plane density}
\label{sec:density}
This statistic uses the non-continuity of electromagnetic showers to tag electrons and positrons. Electromagnetic showers in the EMR have multiple and disconnected tracks. Planes on the path of the shower are often left without any recorded hit. This characteristic can be contrasted with a muon track that uniformly deposits significant amounts of energy along its path.

Plane density, $\rho_P$, is defined as the fraction of planes on the path of the particle that record at least one hit. More specifically, it is the average of that density in the two projections, i.e:

\begin{equation}
\rho_P = \frac{N_x + N_y}{Z_x + Z_y} = \frac N{Z_x + Z_y}\, ;
\end{equation}
with $N_q$ the number of planes hit in the $qz$ projection and $Z_q$ the plane number of the most downstream plane in the $qz$ projection, $q=x,y$. $N$ is the total number of planes hit.

\begin{figure}[tbp]
 \centering
 \includegraphics[width=.85\textwidth]{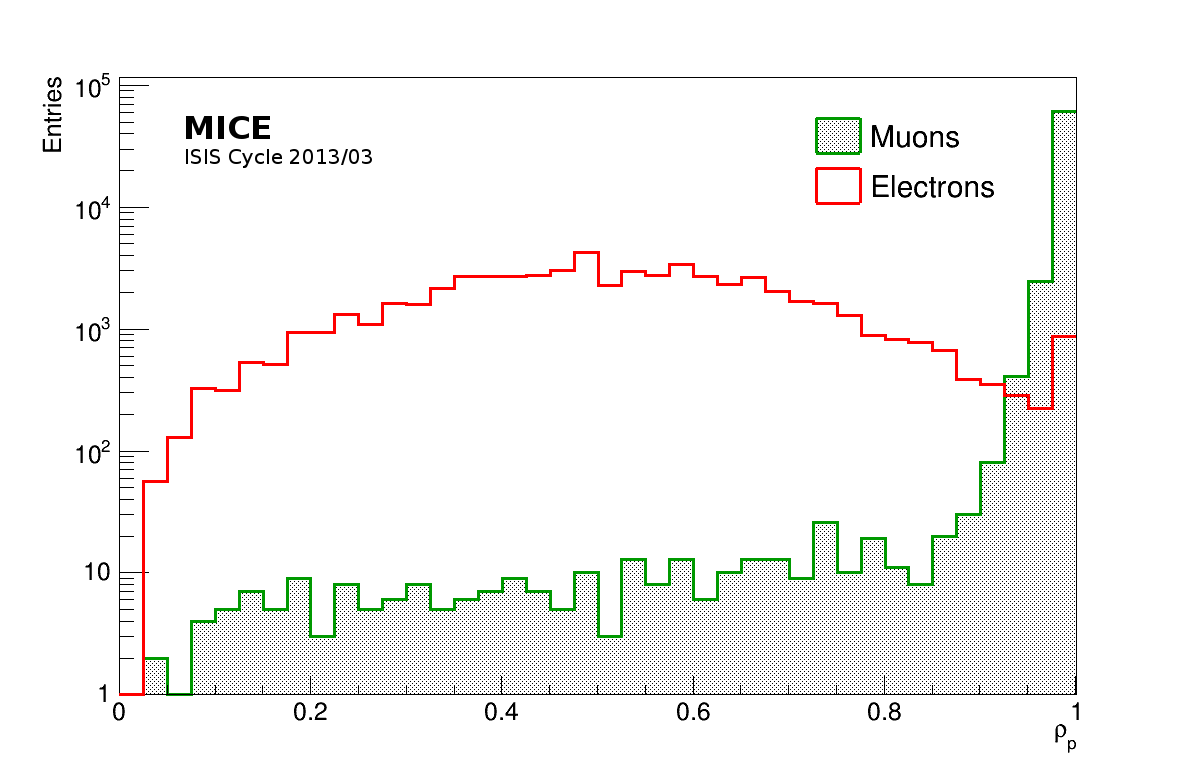}
 \caption[Muon vs electron density]{Logarithmic scale distributions of the plane density ($\rho_P$) for muons and electrons. The integrated electron sample has been normalised to the number of muons.}
 \label{fig:muvse_density}
\end{figure}

The graph in figure~\ref{fig:muvse_density} shows the plane-density distributions for particles tagged as muons and electrons. Muons have densities close to 1 while electrons have a widely spread distribution centred around $\rho_P\sim0.5$. It is striking to see how few electron entries overlap the muon cases.

\subsection{Shower spread}
\label{sec:spread}
In this discriminator, only the most energetic hit in each plane, i.e the bar that records the largest time-over-threshold, is kept. This step allows for the rejection of low-energy crosstalk that would dramatically increase the spread of muon tracks. Crosstalk occurs when some of the light carried by a readout fibre leaks onto adjacent channels on the MAPMT, causing artificial hits in bars up to 13\,cm away from the actual muon track \cite{Drielsma:xt}. Because of the fibre-to-multi-anode PMT mapping, crosstalk only occurs within a single plane, i.e. at a fixed depth in the detector and never between two layers of the calorimeter.

\begin{figure}[tbp]
 \centering
 \includegraphics[width=.85\textwidth]{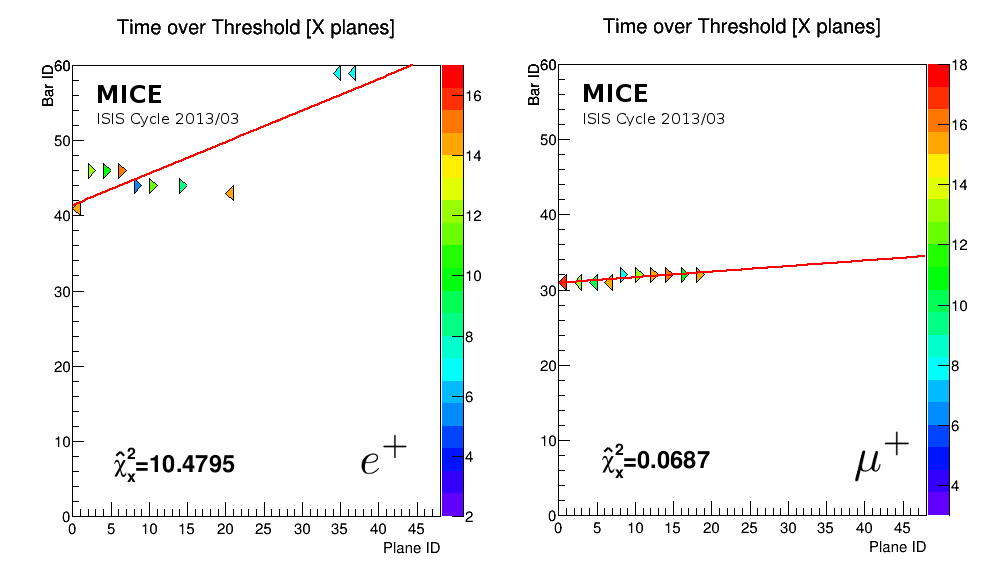}
 \caption[Electron spread]{Fitted electron shower and muon track in the $xz$ projection. The location of a hit is defined by the plane number (Plane ID, 0--47) and bar number (Bar ID, 1--59).}
 \label{fig:fit}
\end{figure}

The remaining $N_q$ hits in a given projection are converted into a set of coordinates $(z_i, q_i)$, $i=1,...,N_q$ and a line is fitted through these coordinates; an example is shown in figure~\ref{fig:fit}.

The trajectories of the particles in the two projections may be parametrised using a function of the form $f_q(z)=a_q z+b_q$. A linear, least-squares fit was performed by minimizing the normalised $\chi^2$ given by:
\begin{equation}
\hat{\chi}_q^2\equiv\chi_q^2/\nu_q = \frac1{\nu_q}\sum_{i=1}^{N_q}\frac{(q_i-f_q(z_i))^2}{\sigma_i^2}=\frac1{\nu_q\sigma_0^2}\sum_{i=1}^{N_q}(q_i-f_q(z_i))^2=\frac1{N_q}\sum_{i=1}^{N_q}(q_i-f_q(z_i))^2\, .
\end{equation}
with $\nu_q = \sum_{i=1}^{N_q}\sigma_i^{-2}$, the normalisation constant. As all the points have the same uncertainty (same bar geometry), it is possible to extract the common $\sigma_0^2\equiv\sigma_i^2$ out of the sum which simplifies the development. Given this definition, it can be shown that the values of $a_q$ and $b_q$ that minimize the normalised $\chi^2$ require that:
\begin{itemize}
\item The barycentre $(\langle z\rangle, \langle q\rangle)$ of the points belongs to the polynomial, i.e $\langle q\rangle = a_q\langle z\rangle + b_q$; and
\item $a_q=V_{zq}/V_z$ where $V_{zq}$ is the covariance of $z$ and $q$ and $V_z$ is the variance of $z$.
\end{itemize}

Under these conditions, the normalised $\chi^2$ in each projection is:
\begin{equation}
\hat{\chi}_q^2 = \sigma_q^2\left(1-\mathrm{corr}^2(z,q)\right)\, ;
\end{equation}
with $\mathrm{corr}(z,q)=V_{zq}/\sigma_z\sigma_q$ and $\sigma_{z,q}$ the standard deviation of the $z$ and $q$ coordinates respectively. The normalised $\chi^2$ is proportional to $\sigma_q^2$, i.e the variance of the $q$ coordinate, which means that the normalised $\chi^2$ increases with the lateral spread of the event in $q$.

\begin{figure}[tbp]
 \centering
 \includegraphics[width=.85\textwidth]{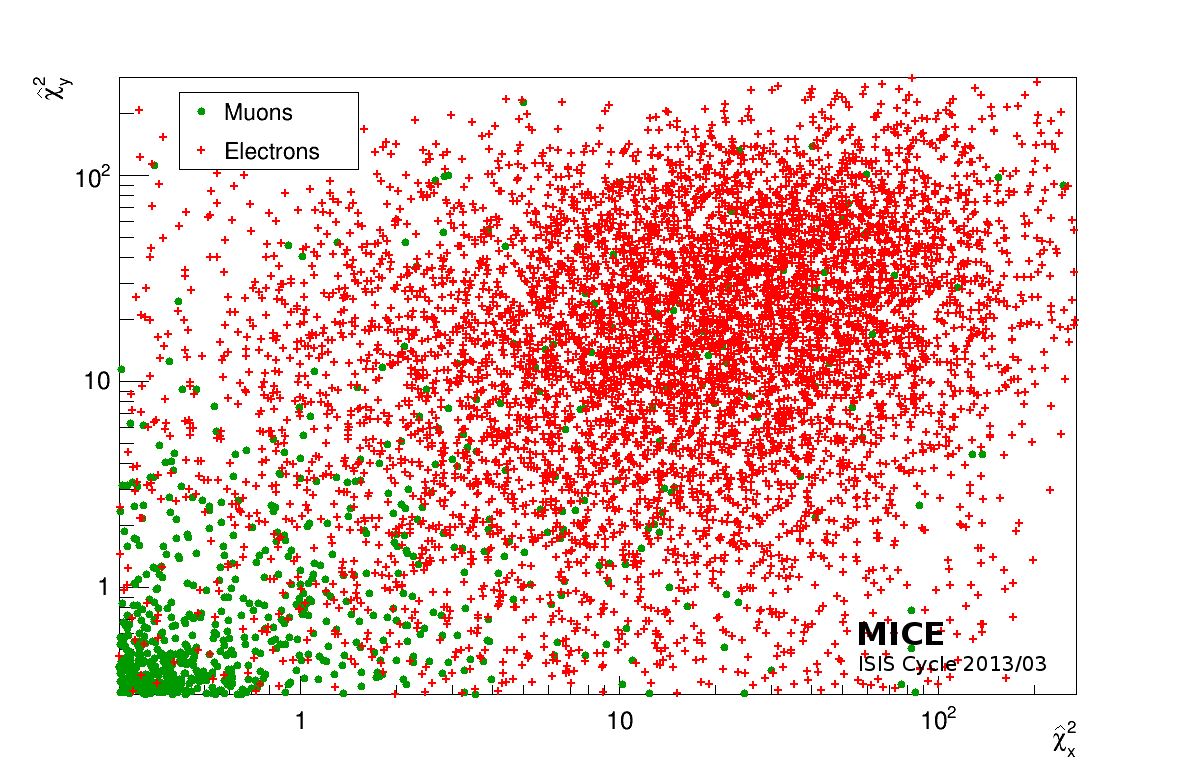}
 \caption[Muon vs electron chi2]{Logarithmic scale scatter plot of the muon and electron samples in the $(\hat{\chi}_x^2, \hat{\chi}_y^2)$ space. $\hat{\chi}_q^2$ is the normalised $\chi^2$ of the linear fit in the $qz$ projection with $z$ along the beam line.}
 \label{fig:muvse_chi2}
\end{figure}

In figure~\ref{fig:muvse_chi2}, $\hat{\chi}_y^2$ is plotted against $\hat{\chi}_x^2$ for particles identified as electrons and muons. As expected from the definition of these quantities, the majority of the muon sample has a small $\chi^2$ in each of the two projections and is located close to the origin. For electrons, the distribution populates the region of large $\hat{\chi}_x^2$ and $\hat{\chi}_y^2$.

% --------------------------------------- %
% ------------TEST STATISTIC------------- %
% --------------------------------------- %
\section{Electron-Muon separation}
The two variables defined in sections \ref{sec:density} and \ref{sec:spread} were used to develop a general test statistic, $T$, by which to distinguish muons from electrons. Given an unknown particle $P$, boundaries, $T_C$, were defined on these variables such that the null hypothesis $H_0$ (\textit{``is a muon''}) or the alternative hypothesis $H_1$ (\textit{``is an electron''}) can be tested.

Two parameters were defined to allow for the hypotheses $H_0$ and $H_1$ to be tested. The loss, $\alpha$, is defined as the proportion of real muons that test negative to $H_0$, i.e. an error of the first kind or false negatives. The contamination, $\beta$, is defined as the proportion of real electrons that test positive to $H_0$, i.e. an error of the second kind or false positives. In this context, the most relevant space to work in is the $(\alpha,\beta)$ space. A test is said to be uniformly more efficient if it has a smaller contamination for any given loss~\cite{james2006statistical}.

When the choice of test has been made, a value of the optimal cut, $T_C^*$, has to be calculated. The optimal value must be such that the corresponding point $(\alpha^*,\beta^*)$ in the $(\alpha,\beta)$ space is as close to the origin as possible. Consider the quantity $\Delta=\sqrt{\alpha^2+\beta^2}$ as the distance between a point of this space to its origin. The value of the cut with the smallest distance $\Delta_{\mathrm{min}}$ is the optimal cut. A perfect test would achieve $\alpha^*=\beta^*=0$, i.e. no overlap between the two samples.

\subsection{Plane density test}
Muons tend to have a much higher plane density than their decay products. Given an unknown particle type $P$, define a cut, $\rho_C$, such that:
\begin{equation}
\rho_P > \rho_C \rightarrow H_0 \nonumber\,;\,\mathrm{or}\,\rho_P \leq \rho_C \rightarrow H_1\, .
\end{equation}

\begin{figure}[tbp]
 \centering
 \includegraphics[width=.85\textwidth]{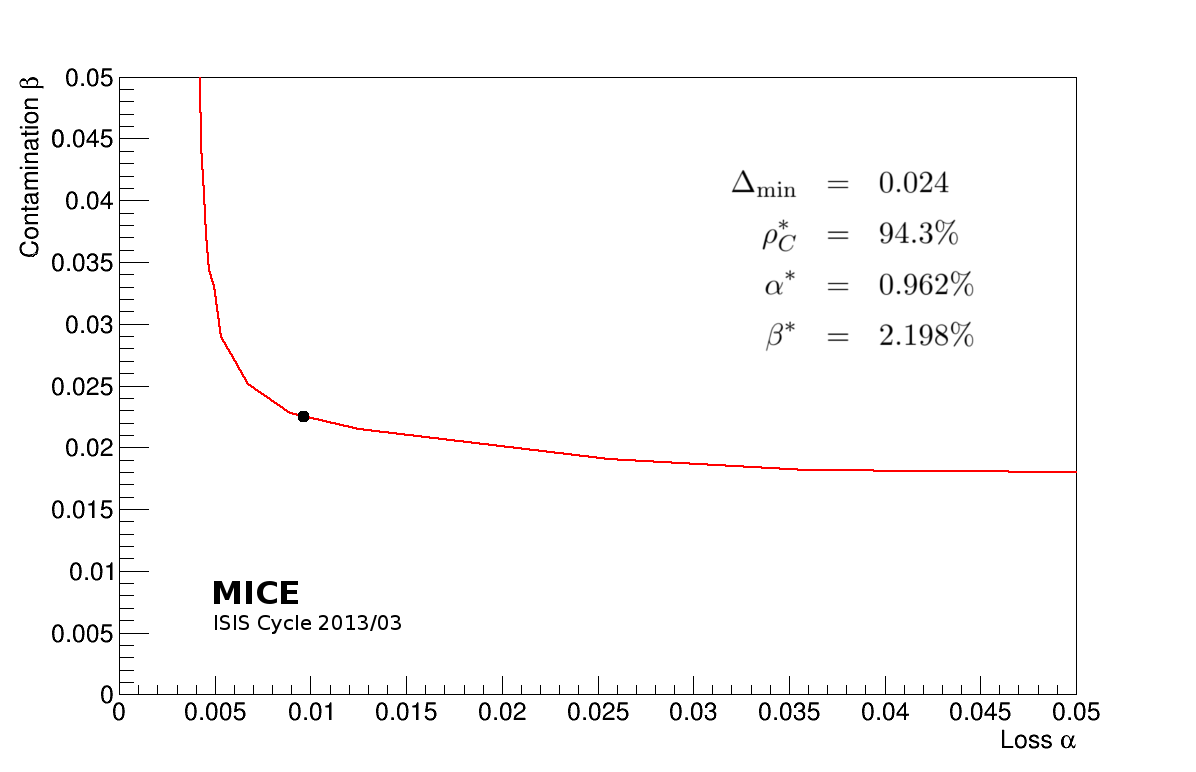}
 \caption[Cuts on density]{Percentage of the electron sample tagged as a muon ($\beta$) as a function of the loss of real muons ($\alpha$) for different values on the cut $\rho_C$. The black dot represents the optimal point of the curve.}
 \label{fig:density_cont}
\end{figure}

The particle is tagged as a muon if its plane density is greater than $\rho_C$ and as an electron otherwise. The muon and electron samples are reprocessed for a wide array of cuts and the values of the contamination and the loss are computed for each of them. The results of this scanning are represented in figure~\ref{fig:density_cont} in the $(\alpha,\beta)$ space. The points occupy a single curve in this space because they are the result of a one dimensional cut. The curve is monotonic as raising the cut increases the loss while reducing the contamination.

The distance, $\Delta$, is calculated for each point and the optimum cut is determined. With this single-variable cut, more than 99\% of the entire muon sample is identified correctly and the electron contamination is $\sim$2.2\% in the muon sample.

\subsection{Spread test}
\begin{figure}[tbp]
 \centering
 \includegraphics[width=.85\textwidth]{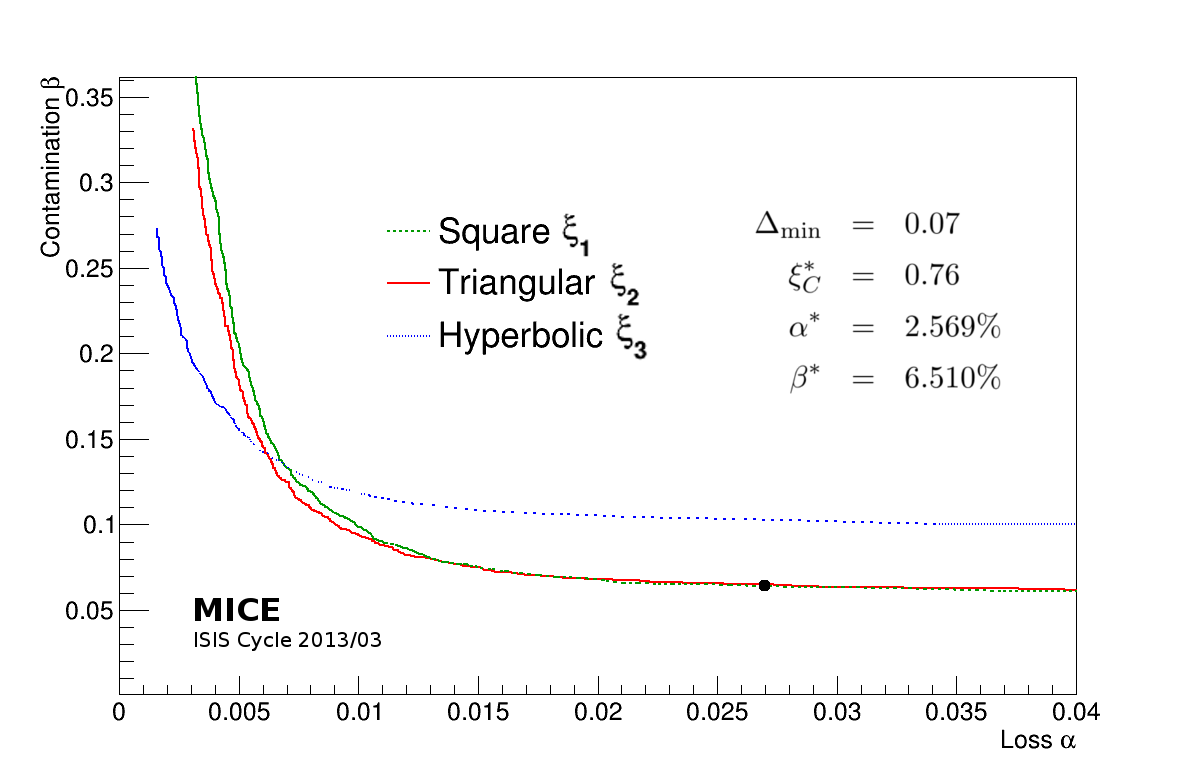}
 \caption[Cuts on chi squares]{Percentage of the electron sample tagged as a muon ($\beta$) as a function of the loss of real muons ($\alpha$) for three choices of test statistic $\xi$. The black dot is the optimal point regardless of the choice of test statistic.}
 \label{fig:chi2_cont}
\end{figure}

The shower spread is summarised by two separate values for the two projections $\hat{\chi}_{x}^2$ and $\hat{\chi}_{y}^2$. The second test statistic, $\xi$, is a combination of these two values. The following tests were considered in this analysis:

\begin{enumerate}
\item $\xi_1 = \max_{q=x,y}{\hat{\chi}_{q}^2}$; largest of the $\hat{\chi}_q^2$. A cut on $\xi_1$ limits a square area in the $(\hat{\chi}_{x}^2,\hat{\chi}_{y}^2)$ space;
\item $\xi_2 = \hat{\chi}_{x}^2 + \hat{\chi}_{y}^2$; sum of the $\hat{\chi}_q^2$. A cut on $\xi_2$ limits a triangular area in the $(\hat{\chi}_{x}^2,\hat{\chi}_{y}^2)$ space;
\item $\xi_3 = \hat{\chi}_{x}^2 \times \hat{\chi}_{y}^2$; product of the $\hat{\chi}_q^2$. A cut on $\xi_3$ limits a hyperbolic area in the $(\hat{\chi}_{x}^2,\hat{\chi}_{y}^2)$ space.
\end{enumerate}
A study of the electron contamination, $\beta$, as a function of the muon loss, $\alpha$, was performed to determine the best choice for $\xi$. The results of this analysis are shown in figure~\ref{fig:chi2_cont}. For losses below $\sim$ 0.5\%, the hyperbolic combination is the most efficient test, but the triangular test yields the lowest contamination for losses above $\sim$ 0.5\% and globally gives the best optimal cut. The square cut produces results similar to the triangular one, but is poorer for low-loss values. Therefore, $\xi\equiv\xi_2=\hat{\chi}_{x}^2 + \hat{\chi}_{y}^2$ was chosen as the test statistic. 

Muons produce very low $\chi^2$ straight tracks in the EMR while electrons, that shower in KL, yield much higher values of $\chi^2$. Given an unknown particle $P$, define a cut, $\xi_C$, such that:
\begin{equation}
\xi < \xi_C \rightarrow H_0 \nonumber\,;\,\mathrm{or}\,\xi \geq \xi_C \rightarrow H_1\, .
\end{equation}
The particle is tagged as a muon if $\xi$ is below a certain threshold and as an electron otherwise. A scan for different $\xi_C$ produces the data in figure~\ref{fig:chi2_cont} in the $(\alpha,\beta)$ space. Once again, for a given choice of test, the points occupy a single dimension and produce a monotonic curve, when the cut is raised the loss decreases and the contamination grows.

The optimal point is obtained by minimizing the distance $\Delta$. The spread test is not as efficient as the density test as both the loss and the contamination are greater. Approximately 97.4\% of the muon sample is identified correctly while the electron contamination of the final sample is $\sim$ 6.5\%. While not optimal in isolation, the spread test can be used in addition to the density test.

\subsection{Multivariate test}
Combining the two test statistics $\rho_P$ and $\xi$ yields better results than either statistic alone. Given an unknown particle species, consider a set of cuts $\rho_C$, $\xi_C$ such that:
\begin{eqnarray}
\centering
\rho_P > \rho_C \cap \xi < \xi_C & \rightarrow & H_0 \nonumber\,;\,\mathrm{or} \\
\rho_P \leq \rho_C \cup \xi \geq \xi_C & \rightarrow & H_1.
\end{eqnarray}
This is a straightforward combination of the two single-variable tests and represents a triangular prism in the $(\rho_P, \hat{\chi}_x^2, \hat{\chi}_y^2)$ space. It has the advantage of including both features of each particle type into one single test. Muon events that are both dense and narrow will be tagged $H_0$ while non-continuous and wide electron events will not.

\begin{figure}[tbp]
 \centering
 \includegraphics[width=.85\textwidth]{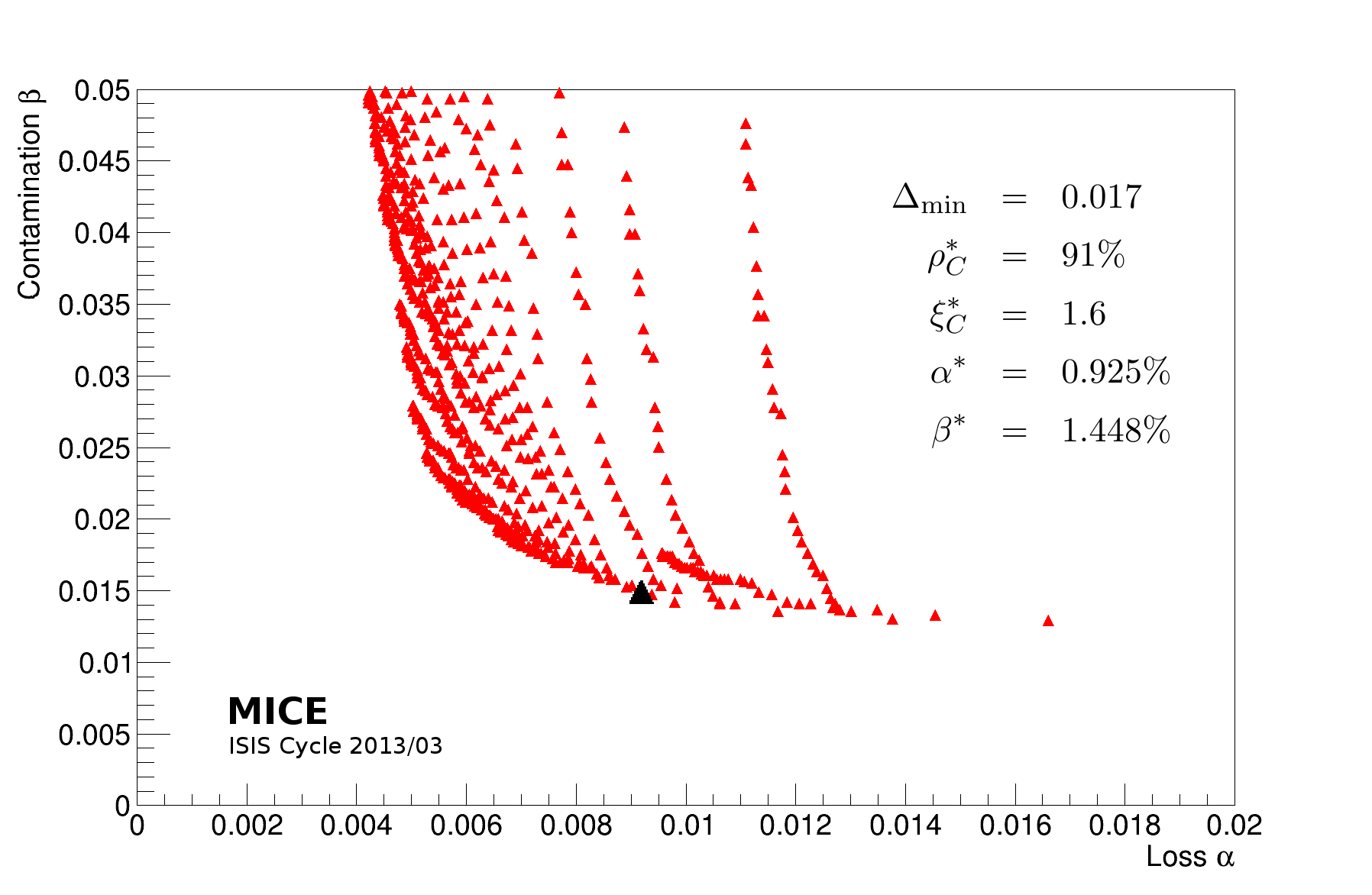}
 \caption[Combined cuts]{Percentage of the electron sample tagged as a muon ($\beta$) versus the loss of real muons ($\alpha$) in the multivariate analysis. The large black triangle is the optimal point.}
 \label{fig:combined_cont}
\end{figure}

A scan of both cuts produces the plot in figure~\ref{fig:combined_cont}. The points do not form a single monotonic function in the $(\alpha, \beta)$ space, showing that the two variables $\rho_P$ and $\xi$ are not entirely correlated, i.e. that their combination carries more information than either one alone. For a given $\rho_C$, the contamination increases with $\xi_C$.

The most critical element of this electron-muon-separation analysis was to determine the optimal point of this multivariate test, as it provides the best separation efficiency by combining single-variable tests. The optimum choice of cuts indicated in figure~\ref{fig:combined_cont} yields a muon loss of $\sim$ 0.9\% and an electron contamination of below 1.45\%. Both figures are lower than the single-variable tests and hence the multivariate analysis is superior to either single-variable analysis.

For instance, in case of equal abundance of muons and electrons, the EMR will deliver a muon-sample purity of 98.55\% with a loss below 1\%. Taking all the beam settings combined, the abundance ratio of electrons to muons was 11.7\%, producing a purity greater than 99.8\% in the final muon sample. In the muon-dominated beam used to demonstrate ionization cooling, the purity provided by the EMR will be much greater than the 99\% required by MICE~\cite{MICE_PID}.

\subsection{Momentum dependence}
The previous sections show that the detector functions as designed and reaches the desired efficiency for samples of muons and electrons. It is necessary to check that the multivariate test developed in this analysis is consistently powerful for all momenta set at D2.

\begin{figure}[tbp]
 \centering
 \includegraphics[width=.85\textwidth]{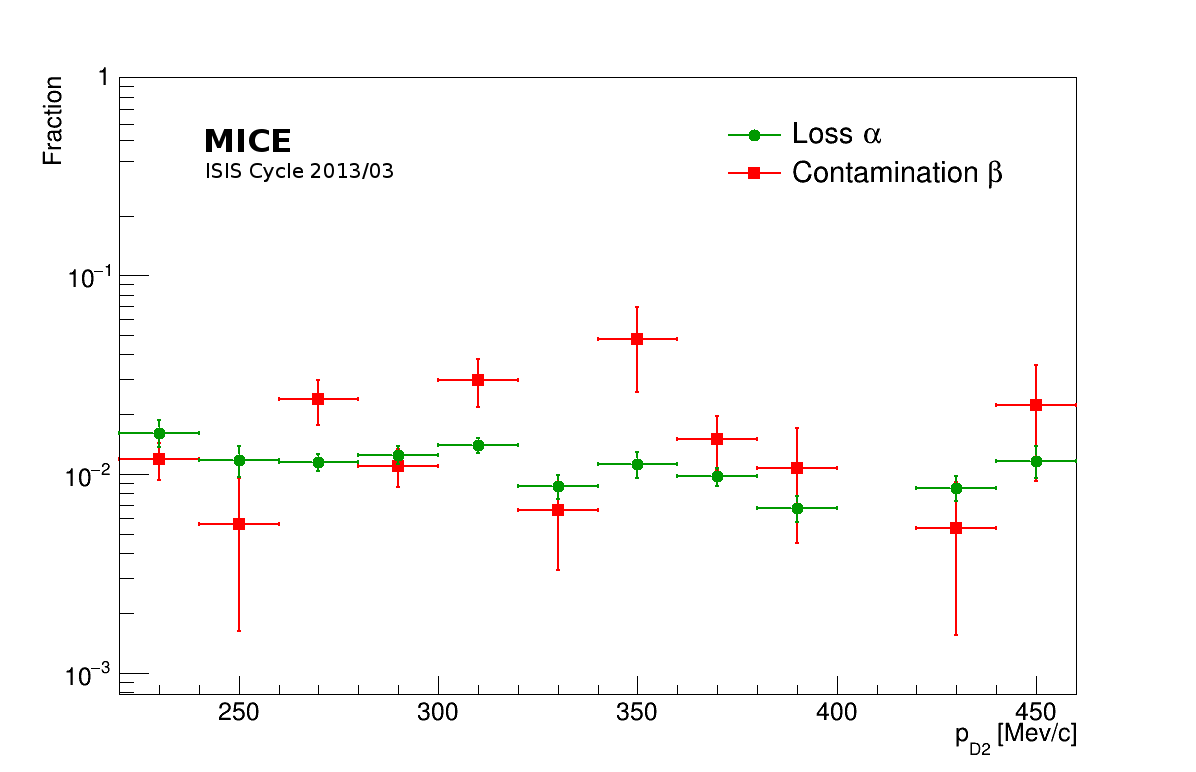}
 \caption[Contamination at different momenta]{Percentage of electron contamination and muon loss for different ranges of momentum set at D2 ($p_\mathrm{D2}$). The error bars are based on the statistical uncertainty in a bin.}
 \label{fig:cont_mom}
\end{figure}

The optimal cuts, $\rho_C^*=0.91$ and $\xi_C^*=1.6$, were applied for different ranges of momentum and the level of contamination and loss was determined for each range. The loss, $\alpha$, in figure~\ref{fig:cont_mom} is $\sim$1\% for the full momentum range. The contamination, $\beta$, fluctuates around 2\%.

% --------------------------------------- %
% --------MOMENTUM RECONSTRUCTION-------- %
% --------------------------------------- %
\section{Muon momentum reconstruction}
In its cooling demonstration, MICE will measure the phase-space reduction of a muon beam traversing the cooling cell. This muon beam is selected upstream of the cooling cell by two dipoles in such a way that a beam with very high muon purity enters the experiment~\cite{BeamlinePaper}. The two trackers measure the muon momentum with great accuracy upstream and downstream of the absorber, provided that the muon has a sufficient transverse momentum, $p_T$. It is not important for the EMR to be able to measure the pion momentum or that of the rejected electrons. It is interesting, however, to estimate how well the EMR can reconstruct the muon momentum.  The EMR can assist the trackers with low $p_T$ muon tracks and can provide a redundant measurement for larger $p_T$.

\subsection{Range reconstruction}
In order to avoid problems induced by MAPMT crosstalk, the most energetic hit in each plane, selected in the same way as was done for the estimation of the spread, was used for the range reconstruction.

\begin{figure}[tbp]
 \centering
 \includegraphics[width=.85\textwidth]{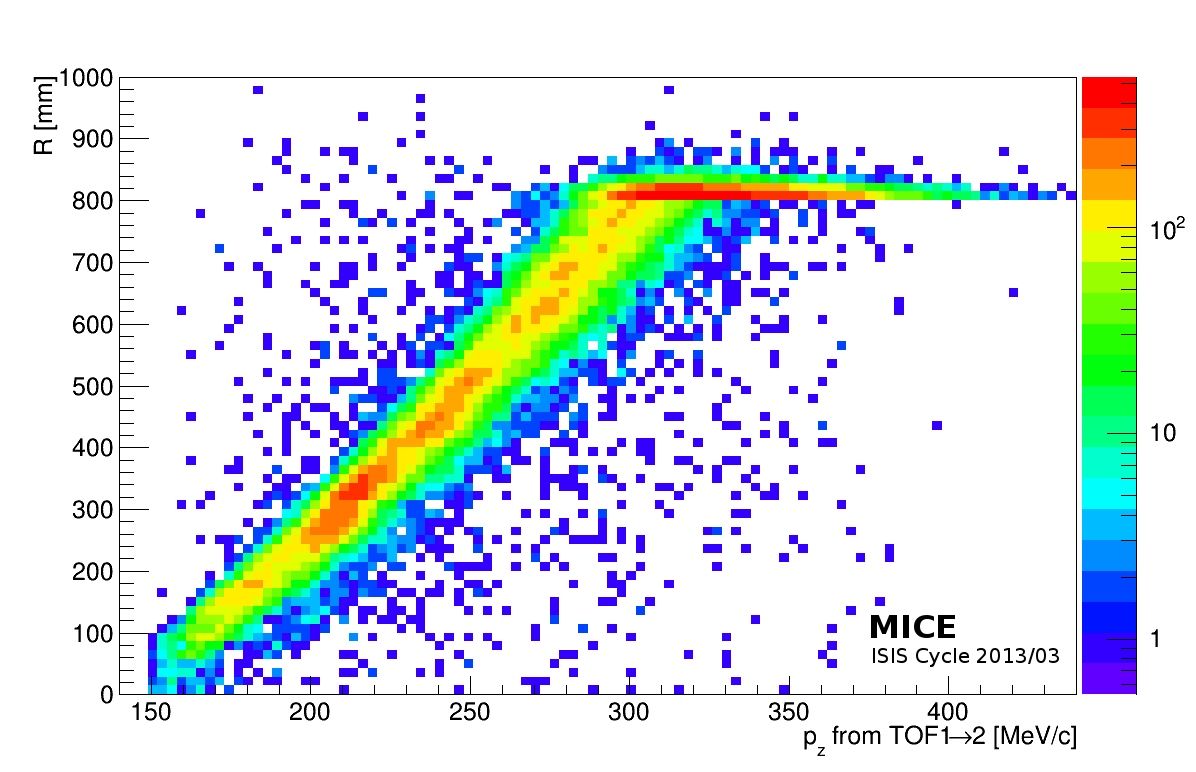}
 \caption[Range of muons vs momentum]{Muon range as a function of the momentum reconstructed from TOF1$\rightarrow$2. At 280\,MeV/$c$ and above, muons can traverse the entire detector without stopping, hence the plateau at $R\sim$ 816\,mm.}
 \label{fig:range_vs_mom}
\end{figure}

The reconstruction method uses the linear fit defined in section~\ref{sec:spread}. The muon track was fitted in its two separate projections and the pitch of its trajectory was measured in each of them. The angles in the $xz$ and $yz$ projections are $\theta_x$ and $\theta_y$, respectively. Provided that the last bar hit by the track has the coordinates $(x_N,y_N,z_N)$, the total range is:
\begin{equation}
R = z_N\sqrt{1+\left(\tan^2\theta_x+\tan^2\theta_y\right)} \equiv z_N/\cos\theta\, ;
\label{eq:range}
\end{equation}
with $\theta = \tan^{-1}\left(\sqrt{\tan^2\theta_x + \tan^2\theta_y}\right)$, the total angle with respect to the $z$ axis. The results are shown for the entire muon sample in figure~\ref{fig:range_vs_mom}.

\subsection{Theoretical approximation}
In order to give an estimate of the muon momentum from a single range measurement it is necessary to develop a theoretical understanding of the dependence of range on momentum. A fit of the data recorded is used to provide a single estimate of the muon momentum downstream of the cooling channel, $p_\mathrm{d}$, as a function of the range in the EMR. We define an invertible function, $f$, such that:
\begin{equation}
R = f(p_\mathrm{d})\iff p_\mathrm{d}=f^{-1}(R)\, .
\end{equation}

To construct a manageable function, the EMR geometry is reduced to a simple 816\,mm thick block of polystyrene as it is the base material used in the plastic scintillator bars. Finding the range of muons in the EMR is equivalent to computing the range in polystyrene. In this situation, the best approach is the same CSDA method used to calculate the energy loss. For an average muon stopping power, $\langle dE/dx \rangle$, of polystyrene, the CSDA range is:
\begin{equation}
R = \int_{E_0}^{0} \frac{dE}{\langle dE/dx\rangle} = \int_{p_0/m_\mu c}^{0} \frac{dp}{\langle dE/dx\rangle}\beta m_\mu c^2\, .
\label{eq:range_th}
\end{equation}

\begin{figure}[tbp]
 \centering
 \includegraphics[width=.85\textwidth]{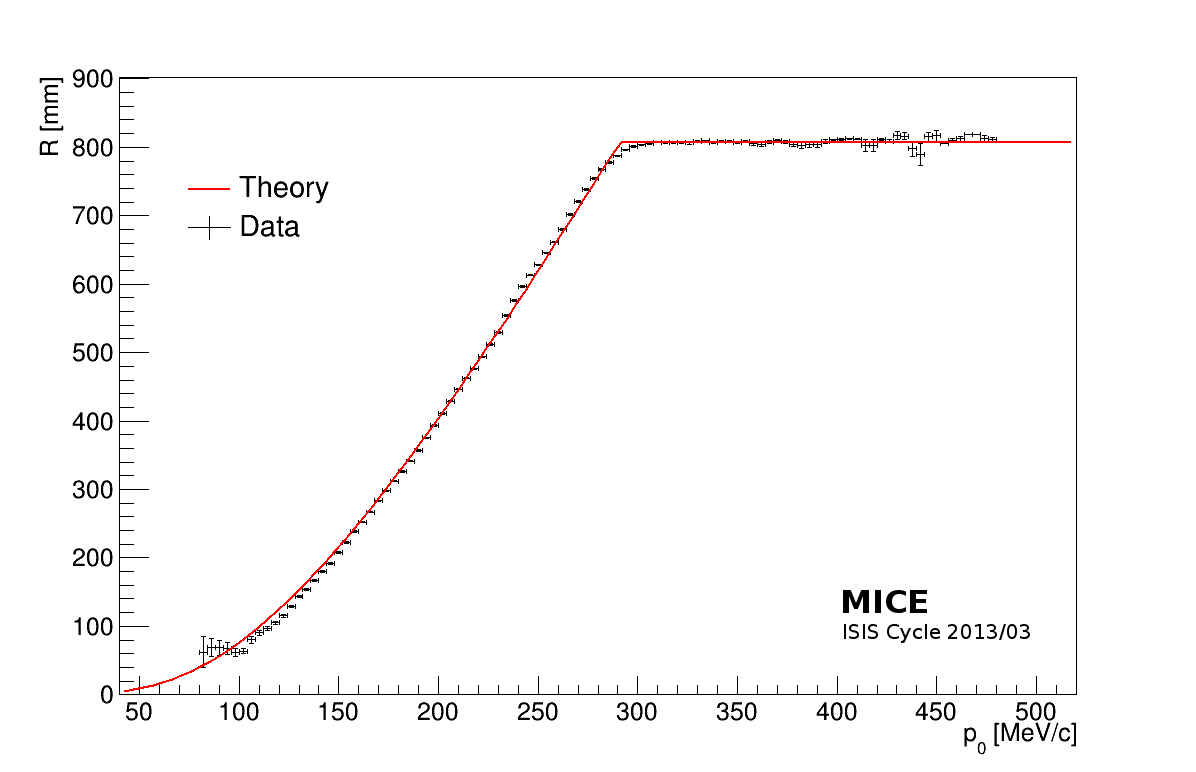}
 \caption[Range of muons data vs theory]{Muon range as a function of the initial momentum $p_0 = p_\mathrm{d} - \Delta p(p_\mathrm{d})$. The error bars represent the bin width in abscissa and the uncertainty of the average reconstructed range in ordinate.}
 \label{fig:range_muon}
\end{figure}

The integral in eq.~(\ref{eq:range_th}) has no analytical solution due to the complexity of the Bethe-Bloch formula. However, it can be integrated numerically for a muon of initial momentum $p_0$. The function can also be inverted numerically to give an estimate of the momentum.

$p_\mathrm{d}$ is the momentum of the muon just before it enters TOF2 and corresponds to the momentum reconstructed from TOF1$\rightarrow$2. This constitutes a complication because it is significantly different from the initial momentum in the EMR, $p_0$. The function $f^{-1}$ provides an estimate of the range for a given $p_0$, but must compensate for the combined momentum loss in TOF2 and KL, $\Delta p(p_\mathrm{d})$. The correction, $p_0 = p_\mathrm{d} - \Delta p(p_\mathrm{d})$, has to be applied muon by muon.

Figure~\ref{fig:range_muon} shows the range of muons in the EMR as a function of their initial momentum, $p_0$. The CSDA prediction shows very good agreement with data for $p_0>150$\,MeV/$c$ while the agreement is acceptable for lower $p_0$ (the plateau of the simulation is added to fit the dimensions of the EMR). The agreement could be improved by a more detailed analysis of the KL, this is beyond the scope of this paper. However, the approximation presented here is a powerful tool that allows an estimate of momentum to be made.

\subsection{Momentum reconstruction accuracy}
Provided with an estimation of the initial momentum, $p_0$, the uncertainty on the unfolding of $p_\mathrm{d}$ may be estimated. There are several sources of uncertainty:

\begin{itemize}
\item The uncertainty on the time of flight $\sigma_t \sim$ 70\,ps and path length $\sigma_s \sim$ 1\,cm;
\item Fluctuation of the momentum-loss in KL as a function of $p_\mathrm{d}$;
\item The uncertainty on the longitudinal position in the EMR, $\sigma_z \sim$ 5\,mm.
\end{itemize}

\begin{figure}[tbp]
 \centering
 \includegraphics[width=.85\textwidth]{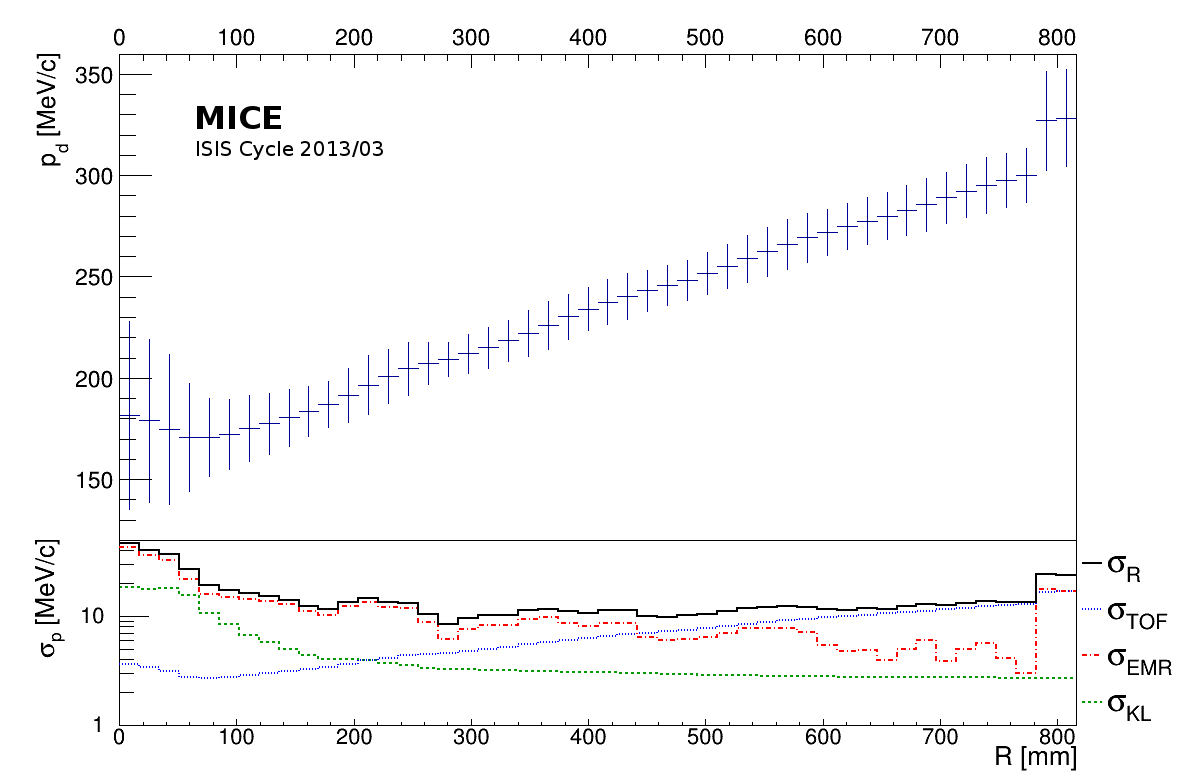}
 \caption[Uncertainty on the range of muons]{Downstream momentum ($p_\mathrm{d}$) as a function of the range ($R$) in the EMR. A bin corresponds to a plane and its error bar represents the RMS of the momentum distribution within the bin.}
 \label{fig:range_uncertainty}
\end{figure}

Figure~\ref{fig:range_uncertainty} shows that the RMS of the momentum distribution in a given plane in the centre of the detector is $\sim$ 10\,MeV/$c$. Each bin represents a single plane of the EMR (17\,mm in depth). At the entrance of the fiducial volume, muons of low momentum stop in the first few planes and the RMS is large. At the back of the detector, the muons with momentum above 320\,MeV/$c$ cross the entire detector and thus the last two planes are biased towards higher values.

The uncertainty on $p_\mathrm{d}$ emerging from the TOF momentum reconstruction, with $t$ the time-of-flight measurement, can be shown to be:
\begin{equation}
\sigma_\mathrm{TOF} = \frac{m_\mu c^2t}{s^2}\left(\left(\frac{ct}{s}\right)^2-1\right)^{-3/2}\sigma_t \oplus \frac{m_\mu c^2t^2}{s^3}\left(\left(\frac{ct}{s}\right)^2-1\right)^{-3/2}\sigma_s\, ,
\label{Eq:tof_sigma}
\end{equation}
with $\oplus$ indicating the quadratic sum. $\sigma_{\mathrm{TOF}}$ ranges from 3--15\,MeV/$c$ for momenta from 150--350\,MeV/$c$. As a bin contains a range of momenta, eq.~(\ref{Eq:tof_sigma}) is averaged over a normal p.d.f. centred on the average momentum in the bin and with a width equal to the bin RMS in order to produce $\sigma_\mathrm{TOF}$.

The spread of the energy loss in KL is computed using the distribution of momentum loss, $f_p(\Delta p)$, given for an array of momenta, $p$, as shown in figure~\ref{KLMomentumLoss}. A single momentum-loss distribution, $f_i(\Delta p)$, constructed for a bin centred in $R_i$, is defined as:
\begin{equation}
f_i(\Delta p) = \int_{0}^{+\infty}f_p(\Delta p)\frac{1}{\sqrt{\pi}\sigma_i}e^{-\frac{(p-p_i)^2}{\sigma_i^2}}\mathrm{d}p\, ;
\end{equation}
with $1/(\sqrt{\pi}\sigma_i) e^{-\frac{(p-p_i)^2}{\sigma_i^2}}$ the normal distribution centred on the average momentum in the bin, $p_i$, and with width equal to the bin RMS, $\sigma_i$. The RMS of $f_i(\Delta p)$ is used as the uncertainty on the momentum from the momentum loss in KL, $\sigma_\mathrm{KL}$.

The total uncertainty on the momentum from the range measurement, $R_\mu$, can be summarized as:
\begin{equation}
\sigma_R=\sigma_{\mathrm{TOF}} \oplus \sigma_{\mathrm{KL}} \oplus \sigma_{\mathrm{EMR}}\, ;
\end{equation}
and therefore, given that the uncertainties are not correlated:
\begin{equation}
\sigma_{\mathrm{EMR}} \sim \sqrt{\sigma_R^2-\sigma_{\mathrm{TOF}}^2-\sigma_{\mathrm{KL}}^2}\, ;
\end{equation}

Figure~\ref{fig:range_uncertainty} shows a compilation of the uncertainties originating from the different components of this analysis. For small values of the range, the EMR has a poor resolution and does not produce reliable momentum reconstruction. After about 10 planes, the total spread reaches its nominal value of $\sim$ 10\,MeV/$c$ and is constant until the last two planes where the values are biased by the muons exiting the detector. The EMR itself, subtracting the errors from KL and TOF, achieves resolutions down to 3\,MeV/$c$ for larger range.

% --------------------------------------- %
% --------------CONCLUSIONS-------------- %
% --------------------------------------- %
\section{Conclusions}
\label{Sect:Conc}
The ability of the Electron-Muon Ranger (EMR) to separate electrons from muons has been demonstrated. It is known from simulation~\cite{Haegel:g4} that muons and electrons produce distinct signal patterns: muons below a certain energy stop in the detector and the energy deposition exhibits a clear Bragg peak, the position of which defines the range. Electrons shower in KL and exhibit multiple tracks in the EMR. In order to verify this, the EMR was exposed to a beam containing both types of particles with momenta in the range 100--400\,MeV/$c$. As expected, electrons and muons produce substantially different patterns so that electrons can be identified with 98.6\% efficiency using the EMR alone.

The detector was not tuned or optimized prior to the measurements. Nevertheless, the detector was able to separate electrons from muons using a simple test statistic on two variables based on the transverse and longitudinal structure of an event.

The detector provides tracking and calorimetric information. Tracks and showers can be reconstructed and identified as muons or electrons. A Bragg peak at the end of muon and pion tracks marks the place where a particle stops and, therefore, helps in measuring the range. It was shown that the muon range can be used to infer the particle momentum in the Continuously Slowing Down Approximation with very good accuracy.

% --------------------------------------- %
% ------------AKNOWLEDGEMENTS------------ %
% --------------------------------------- %
\clearpage
\section*{Acknowledgements}

The work described here was made possible by grants from Department of Energy and National Science Foundation (USA), the Instituto Nazionale di Fisica Nucleare (Italy), the Science and Technology Facilities Council (UK), the European Community under the European Commission Framework Programme 7 (AIDA project, grant agreement no. 262025, TIARA project, grant agreement no. 261905, and EuCARD), the Japan Society for the Promotion of Science and the Swiss National Science Foundation, in the framework of the SCOPES programme. We gratefully acknowledge all sources of support. 

We are also grateful to the staff of the ISIS Department at the Rutherford Appleton Laboratory for the reliable operation of ISIS. We acknowledge the use of Grid computing resources deployed and operated by GridPP in the UK, http://www.gridpp.ac.uk/.

This publication is dedicated to the memory of Mike Zisman who passed away while the data presented here was being prepared for publication.

% --------------------------------------- %
% --------------BIBLIOGRAPHY------------- %
% --------------------------------------- %
\clearpage
\bibliographystyle{utphys}
\bibliography{bibliography}

% --------------------------------------- %
% --------------AUTHOR LIST-------------- %
% --------------------------------------- %
\appendix
\clearpage

\thispagestyle{plain}
\setlength\parindent{0em}%\noindent

{\large \bf The MICE collaboration} \\

\renewcommand{\thefootnote}{\alph{footnote}}
\setcounter{footnote}{0}

{\setlength\parindent{0em}%\noindent

M.~Bogomilov,  R.~Tsenov, G.~Vankova-Kirilova
\\{\it
   Department of Atomic Physics, St.~Kliment Ohridski University of Sofia, Sofia, Bulgaria
}\\

R.~Bertoni, M.~Bonesini, F.~Chignoli, R.~Mazza
\\{\it
Sezione INFN Milano Bicocca, Dipartimento di Fisica G.~Occhialini, Milano, Italy
}\\

V.~Palladino
\\{\it
Sezione INFN Napoli and Dipartimento di Fisica, Universit\`{a} Federico II, Complesso Universitario di Monte S.~Angelo, Napoli, Italy
}\\

A.~de Bari, G.~Cecchet
\\{\it 
Sezione INFN Pavia and Dipartimento di Fisica, Pavia, Italy
}\\

M.~Capponi, A.~Iaciofano, D.~Orestano, F.~Pastore\footnote{Deceased \label{dec}}, L.~Tortora
\\{\it
Sezione INFN Roma Tre e Dipartimento di Fisica, Roma, Italy
}\\

Y.~Kuno, H.~Sakamoto
\\{\it
Osaka University, Graduate School of Science, Department of Physics, Toyonaka, Osaka, Japan
}\\

S.~Ishimoto
\\{\it
High Energy Accelerator Research Organization (KEK), Institute of Particle and Nuclear Studies, Tsukuba, Ibaraki, Japan
}\\

F.~Filthaut\footnote{Also at Radboud University, Nijmegen, The Netherlands}
\\{\it
Nikhef, Amsterdam, The Netherlands
}\\

O.~M.~Hansen, S.~Ramberger, M.~Vretenar
\\{\it
CERN, Geneva, Switzerland
}\\

\begin{alphafootnotes}R.~Asfandiyarov, P.~Bene, A.~Blondel, F.~Cadoux, S.~Debieux, F.~Drielsma\footnote{Corresponding author.}, J.~S.~Graulich, C.~Husi, Y.~Karadzhov, F.~Masciocchi, L.~Nicola, E.~Noah~Messomo, K.~Rothenfusser, R.～Sandstr{\"o}m, H.~Wisting\end{alphafootnotes}
{\it
DPNC, Section de Physique, Universit\'e de Gen\`eve, Geneva, Switzerland
}\\

G.~Charnley, N.~Collomb,  A.~Gallagher, A.~Grant, S.~Griffiths,  T.~Hartnett, B.~Martlew, A.~Moss, A.~Muir, I.~Mullacrane, A.~Oates, P.~Owens, G.~Stokes, P.~Warburton, C.~White
\\{\it
STFC Daresbury Laboratory, Daresbury, Cheshire, UK
}\\

D.~Adams, P.~Barclay, V.~Bayliss, T.~W.~Bradshaw, M.~Courthold, V.~Francis, L.~Fry, T.~Hayler, M.~Hills, A.~Lintern, C.~Macwaters, A.~Nichols, R.~Preece, S.~Ricciardi, C.~Rogers, T.~Stanley, J.~Tarrant, S.~Watson, A.~Wilson
\\{\it
STFC Rutherford Appleton Laboratory, Harwell Oxford, Didcot, UK
}\\

R.~Bayes,  J.~C.~Nugent, F.~J.~P.~Soler
\\{\it
School of Physics and Astronomy, Kelvin Building, The University of Glasgow, Glasgow, UK
}\\

P.~Cooke, R.~Gamet
\\{\it
Department of Physics, University of Liverpool, Liverpool, UK
}\\

A.~Alekou, M.~Apollonio, G.~Barber, D.~Colling, A.~Dobbs, P.~Dornan, C.~Hunt, J-B.~Lagrange, K.~Long, J.~Martyniak, S.~Middleton, J.~Pasternak, E.~Santos, T.~Savidge, M.~A.~Uchida
\\{\it
Department of Physics, Blackett Laboratory, Imperial College London, London, UK
}\\

V.~J.~Blackmore\footnote{Now at Department of Physics, Blackett Laboratory, Imperial College London, London, UK},T.~Carlisle, J.~H.~Cobb, W.~Lau, M.~A.~Rayner, C.~D.~Tunnell
\\{\it
Department of Physics, University of Oxford, Denys Wilkinson Building, Oxford, UK
}\\

C.~N.~Booth, P.~Hodgson, J.~Langlands, R.~Nicholson, E.~Overton, M.~Robinson, P.~J.~Smith
\\{\it
Department of Physics and Astronomy, University of Sheffield, Sheffield, UK
}\\

A.~Dick, K.~Ronald, D.~Speirs, C.~G.~Whyte, A.~Young
\\{\it
Department of Physics, University of Strathclyde, Glasgow, UK
}\\

S.~Boyd,  P.~Franchini, J.~Greis, C.~Pidcott, I.~Taylor
\\{\it
Department of Physics, University of Warwick, Coventry, UK
}\\

R.~Gardener, P.~Kyberd, M.~Littlefield, J.~J.~Nebrensky
\\{\it
Brunel University, Uxbridge, UK
}\\

A.~D.~Bross, T.~Fitzpatrick\footnotemark[1], M.~Leonova, A.~Moretti, D.~Neuffer, M.~Popovic, P.~Rubinov, R.~Rucinski
\\{\it
Fermilab, Batavia, IL, USA
}\\

T.~J.~Roberts
\\{\it
Muons, Inc., Batavia, IL, USA
}\\

D.~Bowring, A.~DeMello, S.~Gourlay, D.~Li, S.~Prestemon, S.~Virostek, M.~Zisman\footnotemark[1]
\\{\it
Lawrence Berkeley National Laboratory, Berkeley, CA, USA
}\\

P.~Hanlet, G.~Kafka, D.~M.~Kaplan, D.~Rajaram, P.~Snopok, Y.~Torun
\\{\it
Illinois Institute of Technology, Chicago, IL, USA
}\\

S.~Blot, Y.~K.~Kim
\\{\it
Enrico Fermi Institute, University of Chicago, Chicago, IL, USA
}\\

U.~Bravar
\\{\it
University of New Hampshire, Durham, NH, USA
}\\

Y.~Onel
\\{\it
Department of Physics and Astronomy, University of Iowa, Iowa City, IA, USA
}\\

L.~M.~Cremaldi, T.~L.~Hart, T.~Luo, D.~A.~Sanders, D.~J.~Summers
\\{\it
University of Mississippi, Oxford, MS, USA
}\\

D.~Cline\footnotemark[1], X.~Yang
\\{\it
University of California, Los Angeles, CA, USA
}\\

L.~Coney, G.~G.~Hanson, C.~Heidt
\\{\it
University of California, Riverside, CA, USA
}\\
}

\end{document}